\documentclass[%
 reprint,
superscriptaddress,
 amsmath,amssymb,
 aps,
]{revtex4-2}

\usepackage{graphicx}% Include figure files
\usepackage{dcolumn}% Align table columns on decimal point
\usepackage{bm}% bold math
\usepackage{float}
\usepackage[utf8]{inputenc}
\usepackage[english]{babel}
\usepackage{xcolor}
\usepackage{siunitx}
\usepackage{amsmath}    % need for subequations
\usepackage{graphicx}   % need for figures
\usepackage{verbatim}   % useful for program listings
\usepackage{color}      % use if color is used in text
\usepackage[T1]{fontenc}
\usepackage[version=3]{mhchem}
\usepackage{siunitx}
\usepackage{tikz}
\usepackage{braket}
\usepackage{caption}
\usepackage{subcaption}
\usepackage{hyperref}   % use for hypertext links, including those to external documents and URLs
\usepackage{xfrac}
\usepackage{babel}
\usepackage{multirow}
\newcommand{\beq}{\begin{equation}\begin{aligned}}
\newcommand{\eeq}{\end{aligned}\end{equation}}

 \usepackage[normalem]{ulem}

\begin{document}

\preprint{}

\title{Two-particle States in One-dimensional Coupled Bose-Hubbard Models}% Force line breaks with \\

\author{Yabo Li}
\affiliation{C. N. Yang Institute for Theoretical Physics and Department of Physics and Astronomy,
State University of New York at Stony Brook, Stony Brook, NY 11794-3840, USA}
\author{Dominik Schneble}
\affiliation{ Department of Physics and Astronomy,
State University of New York at Stony Brook, Stony Brook, NY 11794-3800, USA}
\author{Tzu-Chieh Wei}
\affiliation{C. N. Yang Institute for Theoretical Physics and Department of Physics and Astronomy,
State University of New York at Stony Brook, Stony Brook, NY 11794-3840, USA}
\begin{abstract}
We study dynamically coupled one-dimensional Bose-Hubbard models and solve for the wave functions and energies of two-particle eigenstates.  
Even though the wave functions do not directly follow the form of a Bethe Ansatz,  we describe an intuitive construction to express them as combinations of Choy-Haldane states for models with intra- and inter-species interaction. We find that the two-particle spectrum of the system with generic interactions comprises in general four different continua and three doublon dispersions. 
The existence of doublons depends on the coupling strength $\Omega$ between two species of bosons, and their energies vary with $\Omega$ and interaction strengths.
We give details  on one specific limit, i.e., with infinite interaction, and  derive the spectrum for all types of two-particle states and their spatial and entanglement properties.  We demonstrate the difference in time evolution under different coupling strengths, and examine the relation between the long-time behavior of the system and the doublon dispersion. These dynamics can in principle be observed in cold atoms and might also be simulated by digital quantum computers.
\end{abstract}
\maketitle

\section{Introduction}

The Bose-Hubbard model (BHM), known as one of the simplest models that captures the essence of the superfluid-Mott insulator transition~\cite{Fisher1989}, has given rise to a plethora of studies with cold bosonic atoms in optical lattices~\cite{greiner2002quantum,Lewenstein-Review-2007, BlochDalibardZwerger-2008,Gross-17} in which the ratio between hopping and onsite interaction is widely tunable. Extensions to multi-component BHMs enable studies of polaron physics~\cite{Bruderer2007, Gadway2010} and quantum magnetism ~\cite{Kuklov2003, Altman2003,Jepsen-20,Chung-21}, and by adding a dynamical coupling between the components it furthermore is possible to  implement BHMs that mimic radiative effects~\cite{Vega2008,Navarrete2011,PhysRevA.95.013626,Lanuza2021} in waveguide QED,~\cite{PhysRevA.95.013626,PhysRevA.93.033833,goban2014atom,RevModPhys.87.347,PhysRevA.100.023834,PhysRevA.94.043839, PhysRevResearch.2.043213}
featuring fractional decay, bound states~\cite{PhysRevA.96.023831,krinner2018spontaneous}, and polaritons~\cite{Shi2018Oct,Lanuza2021,Kwon2021}. This also provides a connection to photon-based many-body physics~\cite{Hartmann2006,Greentree2006,Shi2018Oct} in the microwave domain~\cite{astafiev2010resonance,PhysRevLett.107.073601,van2013photon,mlynek2014observation,Ma2019}.

The recent implementation of matter-wave polaritons~\cite{Lanuza2021,Kwon2021} motivates a deeper understanding of the coupled BHM beyond a single excitation.  There have already been some theoretical works on two-particle and multi-particle  waveguide QED and qubit-photon coupled systems using  variational and perturbative methods ~\cite{PhysRevA.93.033833,Shi2018Oct}. In this paper, rather than attempting to solve the full many-body problem, we will give an analytical description of the states with one and two excitations in a one-dimensional coupled Bose-Hubbard model, including the spectrum under different parameters, as well as properties of the states in the continua and of the bound states~\cite{PhysRevLett.64.2418, PhysRevA.95.013626, Shi2018Oct},
The bound states are the so-called \textit{doublon states} \cite{chudnovskiy2012doublon,winkler2006repulsively}, whose wave function is localized in space. In the single-species Bose-Hubbard model, doublon states~\cite{chudnovskiy2012doublon,winkler2006repulsively} exist outside of the two-particle scattering continuum. When the interaction  strength approaches infinity, i.e. $U\rightarrow\infty$, the doublon states form repulsively bound atom pairs that have a well-defined total momentum. 

In this work, we consider the coupled Bose-Hubbard model and provide the complete solution for its two-particle eigenstates. When the intra-species $U\rightarrow\infty$ and the Rabi coupling $\Omega$ is not small (on the scale of the hopping),  
we show below that the doublon states correspond to two particles residing on the same site for different species, but on adjacent sites for the same species, which we will refer to as the `adjacency' feature. When we specify the initial state, by exciting specific empty sites, and letting them evolve, we observe in numerical simulations that the wave function decays incompletely, which is the hallmark of the non-Markovian regime~\cite{krinner2018spontaneous}. When two copies of the same Bose-Hubbard model are coupled together, one of the doublons naturally lies inside a continuum ( different types of states will be explained in detail in Sec.~\ref{sec:spectrum}). This makes it a bound state in the continuum (BIC)~\cite{hsu2016bound,PhysRevLett.122.073601} that is robust in a compact (finite-sized) system. In some other Hubbard models, there are also bound states that can move into and out of the continuum continuously~\cite{2013JPCM...25w5601L, zhang2013bound,della2014floquet}. BICs provide potential applications for quantum memory and some other quantum information processing~\cite{kimble2008quantum,lvovsky2009optical,saglamyurek2018coherent}.

The Bethe Ansatz has been a powerful method to infer the exact wave functions of several systems in 1D, including spin chains, the Fermi Hubbard model, and the Kondo problem, etc.~\cite{essler_frahm_gohmann_klumper_korepin_2005}. In the  Bose-Hubbard model, the exact wave functions of many-body states cannot be inferred from the Bethe Ansatz. However, when the occupation number per site is smaller than 3, this model is solvable. In fact, in several Bose-Hubbard-like models, the two-particle wave functions can be solved. For example, in Refs.~\cite{Valiente_2008, PhysRevA.81.042102, 2009JPhB...42l1001V}, the two-particle states were obtained by solving corresponding Schr\"odinger equations in the infinite system; in Refs~\cite{PhysRevA.102.013510, 2013JPCM...25w5601L}, the Bethe Ansatz was explicitly used.  As explained in details below, in the coupled Bose-Hubbard model, two-particle states can always be written by a Bethe Ansatz   (in the form of a superposition of Choy-Haldane states~\cite{CHOY198049,CHOY198283,PhysRevB.75.115119,PhysRevLett.109.116405}) and solutions to any finite, periodic system can be obtained. Viewed from a different perspective, these few-body states in Bose-Hubbard-like models correspond to 1-particle states in higher-dimensional systems under synthetic dimension mapping~\cite{2021arXiv210410389C}.

Our analysis 
shows that the two-particle spectrum of the system comprises in general four different continua and
three doublon dispersions with generic interactions. Their energies vary with interaction strengths.
We
give details on one specific limit, i.e., with infinite interaction, and analyze the spectrum for all types
of two-particle states and their spatial and entanglement properties. We also study the dynamics of the quantities for an initial simple state under  demonstrate the  time evolution. We observe an interesting relation between the large time scale
behavior of the system and the doublon dispersion.

The remaining structure of the present paper is as follows. In Sec.~\ref{sec:model}, we review the model considered and in Sec.~\ref{sec:single}, we review the solution in the case of single excitations. In Sec.~\ref{sec:two} we first review two-excitations in single-species Bose-Hubbard case and then in Sec.~\ref{sec:mainresults} we generalize the solution to the coupled Bose-Hubbard models without inter-species interaction. The result with inter-species interaction is presented in Appendix~\ref{sec:VI}. In Sec.~\ref{sec:doublon}, we discuss the general properties of the doublon states. They have exponentially decaying wave functions in the thermodynamic limit. In Sec.~\ref{sec:spectrum}, we show the detailed wave function and spectrum in the coupled Bose-Hubbard model and use the inverse participation ratio to demonstrate different localization properties of doublon and scattering states. We also discuss their entanglement between the two species of bosons.  In Sec.~\ref{sec:VII}, we study the time evolution. We find that the existence of doublon states alter the long-term behavior. We make concluding remarks in Sec.~\ref{sec:conclude}.

\section{ The model}
\label{sec:model}
We consider a system of two species of bosons on a lattice with respective hopping strengths $J_1$ and $J_2$   and intra-species interactions $U_1$ and $U_2$ and possibly an energy offset $\Delta$. In addition, there is a direct Rabi coupling $\Omega$ between the two species. This is effectively described by the following coupled   Bose-Hubbard model  (assuming the periodic boundary condition $N+1\equiv 1$ and single-band approximation),
\begin{equation}
\begin{aligned}
 H=&\Delta \sum_j a^{\dagger}_j a_j+\frac{U_1}{2}a^{\dagger}_j a^{\dagger}_j a_j a_j+\frac{U_2}{2} b^{\dagger}_j b^{\dagger}_j b_j b_j
 \\
 &-\sum_{\langle i,j\rangle} (J_1 a^{\dagger}_i a_{j}+J_2 b^{\dagger}_i b_{j})+\Omega\sum_j (a^{\dagger}_j b_j+ \mathrm{h.c.}),
\end{aligned}
\end{equation}
where $i$ and $j$ are the site indices and $\langle i,j\rangle$ indicates nearest-neighbor pair of sites. One can also regard this as two ladders (or copies) of the BH model, with the inter-ladder hopping being $-\Omega$. After Fourier transforming from the position space to momentum space, the Hamiltonian is manifestly conserving the total momentum, 
\begin{equation}
    \begin{aligned}
    H=&\sum_k (\omega_k a^{\dagger}_k a_k+\omega'_k b^{\dagger}_k b_k)+\Omega\sum_k (a^{\dagger}_k b_k +\mathrm{h.c.})
    \\
   +&\sum_{k,p,q} \frac{U_1}{2N}a^{\dagger}_{k-p} a^{\dagger}_{k+p} a_{k-q} a_{k+q} +\frac{U_2}{2N}b^{\dagger}_{k-p} b^{\dagger}_{k+p} b_{k-q} b_{k+q},   
   \end{aligned}
\end{equation}
where $\omega_k=\Delta-2J_1\cos{k}$ and $\omega'_k=-2J_2\cos{k}$ arise from tight binding dispersion relations,  with $k=0, 2\pi/N ,\dots , 2\pi(N-1)/N$ labeling the momentum. Since the total particle number operator $\hat{N}_{\rm excitation}\equiv \sum_k  a^{\dagger}_k a_k + b^\dagger_k b_k$ commutes with $H$, the Hilbert space can be decomposed into sectors differing by the particle number (also referred to as the excitation number): $N_{\rm excitation}=0,1,2,\dots$. 

\section{SINGLE EXCITATIONS}
\label{sec:single}
We will study below eigenstates in the $N_{\rm excitation}=1,2$ subspaces, and note that those in the $N_{\rm excitation}=2$ subspace can be written in the form of Choy-Haldane states~\cite{CHOY198049}, which we review in the next section for a single-species Bose-Hubbard model.

\begin{figure}[h]
\centering
\includegraphics[width=0.8\linewidth]{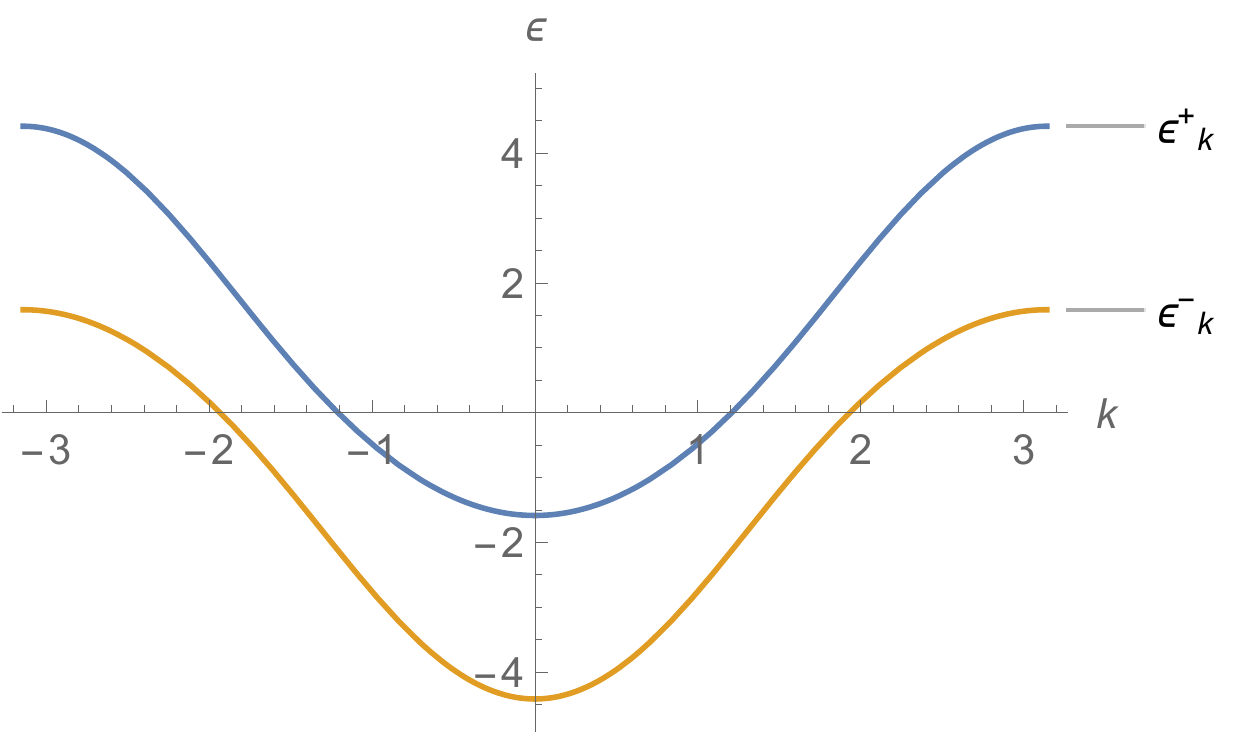}
\caption{The dispersion (energy $\epsilon$ vs. momentum $k$) of the two one-particle  states when $J_1/2=J_2=\Omega$ and $\Delta=0$. The top curve is $\epsilon_k^+$ and the lower curve is $\epsilon_k^-$. 
\label{fig:1D}
}
\end{figure}

 Without loss of generality we write the state of single excitations in the form
\begin{equation}
\ket{\psi}=\sum_{k}(A_k a^{\dagger}_k+B_k b_k^{\dagger})\ket{0}
\end{equation}
and plug it into the
Schr\"odinger equation,
\begin{equation}
    H\ket{\psi}=\epsilon\ket{\psi},
\end{equation}
to obtain
\begin{eqnarray}
&\omega_k A_k + \Omega  B_k =\epsilon A_k, \\
& \Omega A_k + \omega'_k B_k  = \epsilon B_k.
\end{eqnarray}
Then the eigen-energy is the eigenvalue of this $2\times2$ matrix: 
\begin{equation}
\label{eq:1ex}\epsilon_k^\pm=\big(\omega_k+\omega'_k\pm\sqrt{(\omega_k-\omega'_k)^2+4\Omega^2}\,\big)/2.
\end{equation}
We show in Fig.~\ref{fig:1D} an example of the resultant dispersions. When the coupling $\Omega$ is large enough, the spectrum splits into two parts, whose energy does not overlap, and the wave functions are roughly symmetric and antisymmetric between ``$a$'' and ``$b$'' components, respectively. 

Given the solution in momentum space, we could use the Fourier transform to bring the solution back to the position space, such as the components of the wave function, $A_j=\frac{1}{\sqrt{N}}\sum_{k} A_{k}e^{i j k}$, $B_j=\frac{1}{\sqrt{N}}\sum_k B_{k}e^{i j k}$. 
We note that when the context is clear, we use the same symbols $A$ and $B$ in both the position and the momentum spaces. While our analysis above assumes the single-band approximation used in the Bose-Hubbard model, effects of multi-bands can be taken into account, e.g., see Ref.~\cite{Lanuza2021} for the full analytical description of single excitation. We will focus on two-excitation solutions below.

\section{Two-Excitation Solutions}
\label{sec:two}
Having reviewed the single-excitation states, we now turn to two-excitation states. We will first review what was done previously for the single-species Bose-Hubbard model~\cite{CHOY198049,CHOY198283,PhysRevB.75.115119,PhysRevLett.109.116405,Valiente_2008}  and then discuss our results for the coupled case in the next section.
\subsection{Review of Single-species Bose-Hubbard model}

We first recall the Hamiltonian in position space,
\begin{equation}
    H=\Delta \sum_j a^{\dagger}_j a_j-J\sum_j (a^{\dagger}_j a_{j+1} + \mathrm{h.c.}) +\frac{U}{2}\sum_j a^{\dagger}_j a^{\dagger}_j a_j a_j,
\end{equation}
which can be equivalently expressed in momentum space,
\begin{equation}
    H=\sum_k \omega_k a^{\dagger}_k a_k +\frac{U}{2N}\sum_{k,p,q}a^{\dagger}_{k-p} a^{\dagger}_{k+p} a_{k-q} a_{k+q},
\end{equation}
where $\omega_k=\Delta-2J\cos{k}$. We assume the two-excitation wave function to be of the form \begin{equation}
    \ket{\psi}=\sum_{nm} A_{nm} a^{\dagger}_n a^{\dagger}_m \ket{0}=\sum_{pq}A_{pq} a^{\dagger}_p a^{\dagger}_q \ket{0},
\end{equation} 
where $n, m$ are site indices ranging from $0$ to $N-1$, and $p$ and $q$ are momentum variables, ranging from ${2\pi\times 0}/{N}$ to ${2\pi\times (N-1)}/{N}$. Note that we have abused the notation for the coefficients $A$ in both position and momentum representations, as the context will be clear. In the momentum space, since the total momentum is conserved, we can denote it by $P$. Then the only non-zero components are those with momentum indices satisfying $p+q=P$. We shall thus abbreviate $A_{p,P-p}$ as $A_p^{(P)}$ (or even $A_p$) for simplicity in the following. Then similar to the one-excitation case, the Schr\"odinger equation gives
\begin{equation}
     A_{p} = \frac{U}{N} \frac{A^{(P)}}{\epsilon-\omega_p-\omega_{P-p}},
     \label{ms1}
\end{equation}
where $A^{(P)} \equiv \sum_{k}A_{k,P-k}$ and  $\epsilon$ is the eigen-energy.

The position wave function $A_{n,m}$ is a symmetric matrix, and we can write the position-space Schr\"odinger equations as a matrix equation, 
\beq
\epsilon A+TA+AT-U D_A =0,
\label{bhsce}
\eeq
where the hopping matrix $T_{nm}=J(\delta_{n,m-1 \mathrm{mod} N}+\delta_{n,m+1 \mathrm{mod} N})$, and $D_A$ is the diagonal part of $A$, representing double occupancy of a site. The solutions of this matrix equation are of the Choy-Haldane type~\cite{CHOY198049,CHOY198283,PhysRevB.75.115119,PhysRevLett.109.116405},
\begin{equation}
A_{n,m}=\left\{
\begin{aligned}
&e^{i k n+i q m}+s_{k,q}e^{i q n+i k m}, & n\leq m
\\
&A_{m,n},  &n>m.
\end{aligned}
\right.
\label{hc1}
\end{equation}
where 
\begin{equation}s_{k,q}\equiv {[2J(\sin{k}-\sin{q})-iU]}/{[2J(\sin{k}-\sin{q})+iU]}.
\end{equation}
One can verify that this is a solution by substituting this form $A_{n,m}$ into the matrix equation~(\ref{bhsce}) above.
For the periodic boundary condition, the quasi-momentum $k$ (also known as the Bethe parameter) satisfies
\begin{equation}
    e^{-i k N}=\frac{2J(\sin{k}-\sin{q})-iU}{2J(\sin{k}-\sin{q})+iU},
    \label{betheeq}
\end{equation}
and $q=P-k$ from the momentum conservation. Note that from this constraint we know that for a generic $U\ne 0$, the quasi-momentum $k$ is not a well-defined momentum on the lattice, i.e., not a physical momentum, with the latter being ($\frac{2\pi}{N}\times \mathrm{Integer}$). The wave function (\ref{ms1}) in space should be related to that (\ref{hc1}) in momentum by the Fourier transform: $A_{n,m}=\frac{1}{N}\sum_{p,P}A_p^{(P)}e^{ink+im(P-k)}$.

\begin{figure}[h]
     \centering
     \begin{subfigure}[b]{0.8\linewidth}
        \centering
         \includegraphics[width=\linewidth]{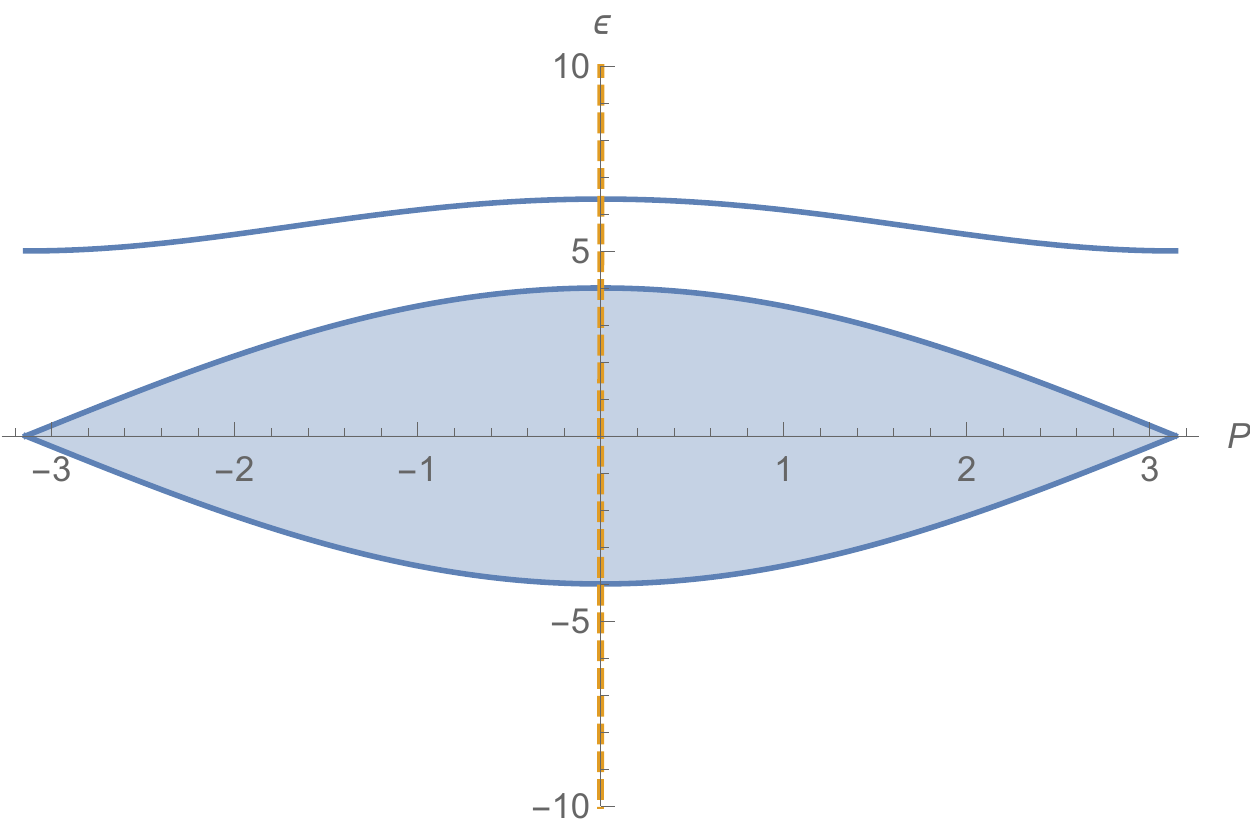}
         \caption{Dispersion of the continuum band and doublon for $U/J=5$. The yellow dashed line indicates the spectrum with total momentum $P=0$,which contains a continuum  in the shaded region (formed by the band of two-particle scattering states) and a doublon state outside of the continuum.}
         \label{bh_dispersion}
     \end{subfigure}
    \hfill
     \begin{subfigure}[b]{0.8\linewidth}
         \centering
         \includegraphics[width=\linewidth]{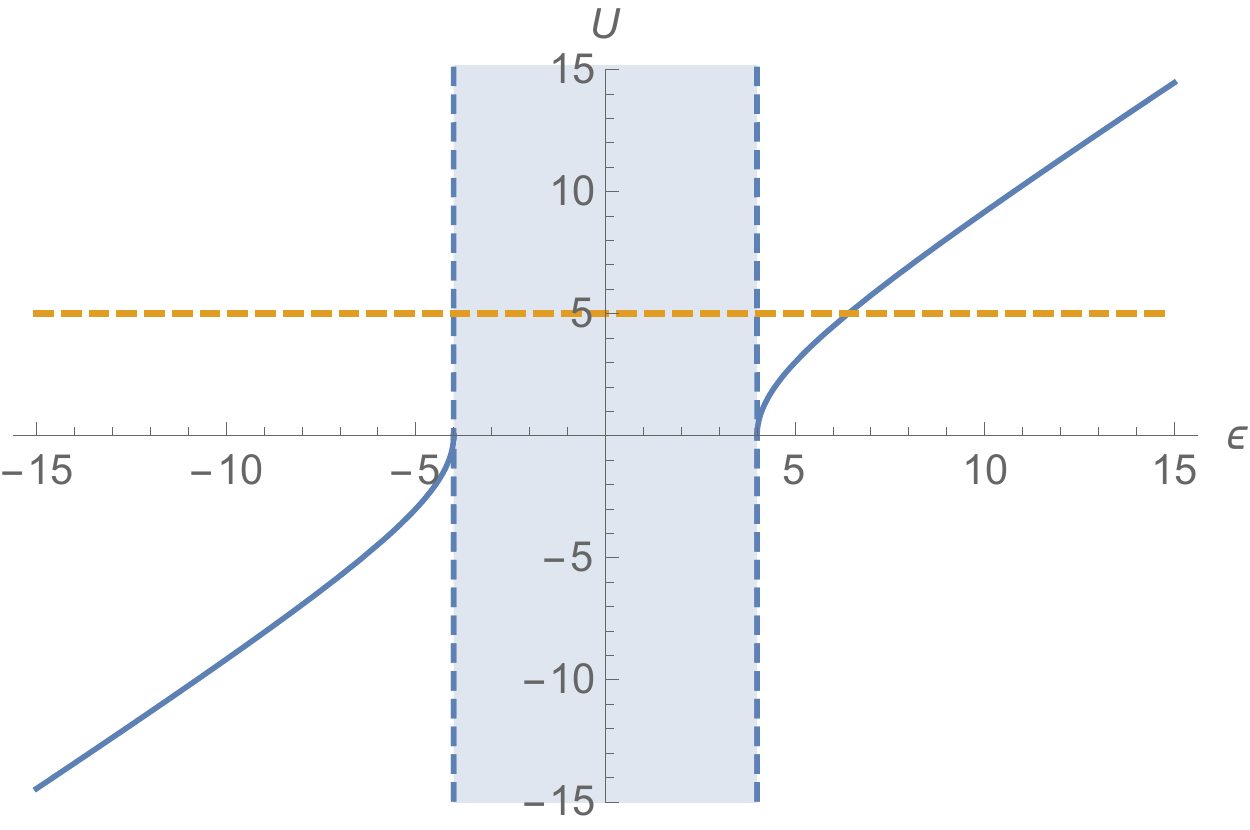}
\caption{The interaction dependence of spectrum with total momentum $P=0$. The yellow dashed line shows the spectrum for a fixed interaction strength $U/J=5$, the horizontal section is the same as the corresponding vertical section in figure~\ref{bh_dispersion}.}
\label{bh_spect}
     \end{subfigure}
     \caption{The interaction and total momentum dependent spectrum of single-species Bose-Hubbard model for two-excitation states in the thermodynamic limit. We have taken  $\Delta=0$, $J=1$. }
\end{figure}

Eq.~(\ref{betheeq}) is referred to as the Bethe equation~\cite{essler_frahm_gohmann_klumper_korepin_2005}. Given parameters $J$ \& $U$ and the total momentum $P$, there are $N-1$ real solutions in the continuum and one complex solution outside of the continuum. The latter, for $U>0$, is  a repulsive  bound pair state (attractive for $U<0$), in which two particles are located at same sites. The whole two-excitation spectrum for fixed $U/J=5$ is shown in Fig.~\ref{bh_dispersion}, where one can see the continuum (i.e. a band) and a separate dispersion curve above.  
 To illustrate the dependence of the spectrum as the interaction $U$ changes, in Fig.~\ref{bh_spect} we display the energy at total momentum $P=0$ horizontally and vary $U$ vertically. This form of the interaction-dependent zero-momentum diagram will be used extensively below. 

The state outside of the continuum (whose dispersion represented by the curve above the continuum band) is called a doublon state~\cite{chudnovskiy2012doublon}, which is a two-particle bound state (in the relative coordinate). In the case of repulsive interaction, it may seem counter-intuitive to have a stable bound state of high energy, but one can understand its existence intuitively as the pair is unable to decay by converting the potential energy into the kinetic energy~\cite{winkler2006repulsively}.  In terms of the mathematics, the corresponding quasi-momentum of a doublon state is complex, reflecting the nature of bound states, whose expression is $k=\frac{P}{2}+\pi-i K$ if $U>0$, and $k=\frac{P}{2}-i K$ if $U<0$, where the imaginary part $K>0$ is determined by the Bethe equation. When $N\rightarrow\infty$, i.e. in the thermodynamic limit, the left hand side of the Bethe equation approaches 0. Thus the equation becomes $2J\cos{\frac{P}{2}}\sinh{K}=U$. The wave function of the doublon state is thus
\begin{equation}
    A_{n,m}=\left\{
    \begin{aligned}
    &e^{-K|n-m|}e^{i(\frac{P}{2}+\pi)(n+m)}, &\mathrm{when}~ U>0,
\\
    &e^{-K|n-m|}e^{i(\frac{P}{2})(n+m)}, &\mathrm{when }~ U<0.
\end{aligned}
\right.
\end{equation} 
We can clearly see that bound-state feature of the exponential decay in the relative position of the two particles, i.e. $e^{-K|n-m|}$. We note that an alternative approach to the one described above~\cite{CHOY198049,CHOY198283,PhysRevB.75.115119,PhysRevLett.109.116405} was also given in Ref.~\cite{winkler2006repulsively,Valiente_2008} by solving the scattering problem for two bosons.

\section{Coupled Bose-Hubbard model without inter-species interaction}
\label{sec:mainresults}
Having reviewed the single Bose-Hubbard model case, we now present our main results for the doubly occupied, coupled Bose-Hubbard model, where $\Omega$ is the coupling strength. In the momentum space, we can assume that the wave function takes the form 
\beq
\ket{\psi}=\sum_{p,q} (A_{p,q}a_p^{\dagger}a_q^{\dagger}+B_{p,q}a_p^{\dagger}b_q^{\dagger}+C_{p,q}b_p^{\dagger}b_q^{\dagger})\ket{0}.
\eeq
Then the corresponding Schr\"odinger equations are
\begin{equation}
\left\{
\begin{aligned}
(\epsilon-\omega_p-\omega_q) A_{p,q}&= \frac{U_1}{N}\sum_{p+q=P}A_{p,q}+\frac{\Omega}{2}(B_{p,q}+B_{q,p}),
\\
(\epsilon-\omega_p-\omega'_{q}) B_{p,q}&=2\Omega A_{p,q} + 2\Omega C_{p,q},
\\
(\epsilon-\omega'_p-\omega'_q)C_{p,q}&=\frac{U_2}{N}\sum_{p+q=P}C_{p,q}+\frac{\Omega}{2}(B_{p,q}+B_{q,p}).
\label{scheoms}
\end{aligned}
\right.
\end{equation}
Since $\sum_{p+q=P}A_{p,q}$ and $\sum_{p+q=P}C_{p,q}$ are constants after fixing the total momentum $P$, we simply denote them by $A$ and $C$, respectively. Solving for $A_p\equiv A_{p,P-p}$ and $C_p\equiv C_{p,P-p}$, we arrive at
\begin{widetext}
\begin{eqnarray}
\label{Ap}
A_p&=\frac{1}{(\epsilon-\omega_p-\omega_q-\eta_{pq})(\epsilon-\omega'_p-\omega'_q-\eta_{pq})-\eta_{pq}^2} \Big( (\epsilon-\omega'_p-\omega'_q-\eta_{pq})\frac{U_1 A}{N}+\eta_{pq}\frac{U_2 C}{N}\Big), \label{eq:Ap}
\\
\label{Cp}
C_p&=\frac{1}{(\epsilon-\omega_p-\omega_q-\eta_{pq})(\epsilon-\omega'_p-\omega'_q-\eta_{pq})-\eta_{pq}^2} \Big( \eta_{pq}\frac{U_1 A}{N}+(\epsilon-\omega_p-\omega_q-\eta_{pq})\frac{U_2 C}{N}\Big),
%\label{ac}
\end{eqnarray}
\end{widetext}
where $\eta_{pq}\equiv\Omega^2(\frac{1}{\epsilon-\omega_p -\omega'_q}+\frac{1}{\epsilon-\omega_q -\omega'_p})$ and $q=P-p$. If we further sum over the momentum index $p$ in the above equations, we would have two linear homogeneous equations for $A$ and $C$. To have a nontrivial wave function the determinant of the two-by-two matrix should be zero. From this we derive the energy for every given value of the total momentum $P$. This allows us to solve for the two-excitation spectrum. We leave some of the details in Appendix~\ref{app:momentum}. In the following, we only use the momentum space result, such as in Eqs.~(\ref{Ap}) and~(\ref{Cp}) to guide  our main approach   that generalizes the real-space Choy-Haldane solution from the single-species case, which is in some sense a Bethe Ansatz approach.

\subsection{Two-species  BH model with coupling $\Omega$}
\label{sec:IVA}
Let us take the simplest case, i.e., $J_1=J_2=J$, $\Delta=0$, $U_1=U_2=U$. Then we have two identical single-body spectra: $\omega_p=\omega'_p=-2J\cos{p}$. The $2\times2$ coefficient matrix mentioned in the last paragraph (see also Eq.~(\ref{ac}) in the Appendix and discussions there) has eigenvectors as $A=-C$ and $A=C$. In the former case, $B_{p,q}=0$, and the equation for $A_{p,q}$ (or $C_{p,q}=-A_{p,q}$) reduces to the  Schr\"odinger equation for the (single-species) Bose-Hubbard model,
\begin{equation}
\label{eqn:mHC}
    A_{p,q}=\frac{1}{\epsilon-\omega_p-\omega_q}\frac{UA}{N},
\end{equation}
which we have seen in Eq.~(\ref{ms1}).
Therefore in this case we have just a single Choy-Haldane state and the solution for $A$ is identical. 

In the case when $A_{p,q}=C_{p,q}$, $B_{p,q}\neq0$ and we should not expect the wave function to be just a single Choy-Haldane state. Observing that, when $A=C$, equation~(\ref{eq:Ap}) for $A$ can be rewritten as
\beq
    A_p&=\frac{\epsilon-\omega_p-\omega_q}{(\epsilon-\omega_p-\omega_q-\eta_{pq})(\epsilon-\omega_p-\omega_q-\eta_{pq})-\eta_{pq}^2}\frac{U A}{N}
    \\
    &=\frac{1}{2}\Big(\frac{1}{\epsilon-\omega_p-\omega_q-2\Omega}+\frac{1}{\epsilon-\omega_p-\omega_q+2\Omega}\Big)\frac{UA}{N}.
\eeq
It is useful to note that the two terms in $A_p$ are of similar form to  Eq.~(\ref{eqn:mHC}), i.e.  the momentum representation of a Choy-Haldane state, except having an extra constant $\pm2\Omega$ in the denominator. Since $\pm2\Omega$ is a constant, we can absorb it separately into $\epsilon$ in the respective denominator. Therefore, we expect that in the case when $A_{p,q}=C_{p,q}$, the solution is a superposition of two Choy-Haldane states, in which $ \epsilon $  is replaced by $\epsilon+2\Omega$ and $\epsilon-2\Omega$, respectively, and we shall write this symbolically as $\ket{\psi}=\ket{HC_1}+\lambda\ket{HC_2}$, where $\lambda$ denotes the relative weight to be determined. 

From the previous section, we know that a single Choy-Haldane state is characterized by the scattering factor $s_{k,q}$. One expects  the two Choy-Haldane states here to have their own respective (fictitious) ``interaction strengths'', which determines the factor $s_i=[{2J(\sin{k}-\sin{q})-i\Tilde{U}_i}]/[{2J(\sin{k}-\sin{q})+i\Tilde{U}_i}]$. (Notice that we put a tilde on $U_i$  to indicate that it is \textit{not} the physical interaction strength in the Hamiltonian.) However, would the conjectured superposition be a consistent solution as $\Tilde{U}_i$'s are not the physical interaction $U$? In other words, can $\Tilde{U}_i$'s be consistently determined from the system's parameters, such as the physical interaction strength $U$, the total energy $\epsilon$, and the total momentum $P$? 

It turns out that not only this superposition trick works, but we also  can determine $U_i$'s in terms of physical parameters. In this and the following examples, we even obtain  simple relations between $\tilde{U}_i$ and the physical interaction $U$.

The way to determine the two fictitious interaction strengths $\Tilde{U}_1$ and $\Tilde{U}_2$ begins as follows. From the energy equation for the two Choy-Haldane states, 
\begin{eqnarray}
    \epsilon-2\Omega&=\omega_{k_1}+\omega_{P-k_1},
\label{eqn:energy1}
    \\
    \epsilon+2\Omega&=\omega_{k_2}+\omega_{P-k_2},
\label{eqn:energy2}
\end{eqnarray}
where $k_1$ and $k_2$ are the quasi-momentum in $\ket{HC_1}$ and $\ket{HC_2}$, respectively, we have the Bethe equations satisfied by these quasi-momenta,
\beq
    e^{-i k_i N}=\frac{2J\big(\sin{k_i}-\sin{(P-k_i)}\big)-i\Tilde{U}_i}{2J\big(\sin{k_i}-\sin{(P-k_i)}\big)+i\Tilde{U}_i}\equiv s_i,
\label{eqn:bethe}
\eeq
for $i=1,2$. Furthermore, making use of the Schr\"odinger equation for $B_{n,m}$, we find that $\Tilde{U}_i$'s are related to $U$ via 
\begin{equation}
\label{eqn:UU12}
 U={2\Tilde{U}_1\Tilde{U}_2}/({\Tilde{U}_1+\Tilde{U}_2}).  
\end{equation} From the three equations above, we finally obtain the relation of the energy $\epsilon$ with the interaction strength $U$ and the total momentum $P$.

We now elaborate on the steps to obtain Eq.~(\ref{eqn:UU12}). To do this, we use the Schr\"odinger equations in the position space, and when $J_1=J_2$, they are
\begin{equation}
\label{eqn:J1=J2}
    \left\{
    \begin{aligned}
        & \epsilon A + T A + A T -U D_A = \Omega B_s,
\\
& \epsilon B+ T B +B T=2\Omega (A+C),
\\
& \epsilon C + T C + C T -U D_C= \Omega B_s,
    \end{aligned}
    \right.
\end{equation}
where $B_s\equiv(B+B^T)/2$ is the symmetric part of $B$ and we also define $B_a\equiv(B-B^T)/2$ to be the antisymmetric part of $B$. Note that the normalization of our wave function is 
\beq
\sum_{n,m} 2|A_{n,m}|^2+ |B_{n,m}|^2+2|C_{n,m}|^2=1.
\label{eq:normalization}
\eeq
When $A=C$, the three equations reduce to two:
\begin{equation}
\left\{
    \begin{aligned}
&\epsilon A + T A + A T -U D_A = \Omega B_s,
\\
&\epsilon B+ T B +B T=4\Omega A.
    \end{aligned}
    \right.
\end{equation}

 Let us  assume $B_a=0$ for now (we will come back to the scenario $B_a\ne0$ below), then we have $B_s=B$. Taking $A$ to be a superposition of two Choy-Haldane states with different weights, $A=\psi_1+\lambda \psi_2$, with these two states $\psi_1$ and $\psi_2$ (in a matrix form) each satisfying an equation similar to the single-species case in Eq.~(\ref{bhsce}),
\begin{equation}
    \left\{
    \begin{aligned}
        (\epsilon-2\Omega) \psi_1 + T \psi_1 +\psi_1 T -\Tilde{U}_1 D_{\psi_1} &= 0,
\\
(\epsilon+2\Omega) \psi_2 + T \psi_2 +\psi_2 T -\Tilde{U}_2 D_{\psi_2} &= 0,
    \end{aligned}
    \right.
\end{equation}
where, for convenience, $D_{\psi_1}$ and $D_{\psi_2}$ are used to denote the diagonal parts of $\psi_1$ and $\psi_2$, respectively.
Notice that {\it if} 
\beq
\Tilde{U}_1(1+s_1)+\lambda\Tilde{U}_2(1+s_2)=U(1+s_1)+\lambda U(1+s_2),
\label{diagonal}
\eeq
 then, according to the first Schr\"odinger equation, $B=2(\psi_1-\lambda\psi_2)$. With this,  we can use the other equation to obtain another relation for the parameter $\lambda$ and the $\tilde{U}_i$'s: 
\beq
\Tilde{U}_1(1+s_1)-\lambda\Tilde{U}_2(1+s_2)=0.
\eeq
Solving the above two equations we obtain the relation $U={2\Tilde{U}_1\Tilde{U}_2}/({\Tilde{U}_1+\Tilde{U}_2})$.

To conclude the above calculations, we have shown that the matrices for the wave function,  $A$ and $C$ (as well as $B$, if nonzero), can be written as combinations of two Choy-Haldane states, each with a fictitious interaction strength $\tilde{U}_i$. Since the diagonal part of the Choy-Haldane state is $D_{nm}=(1+s) e^{iPn}\delta_{nm}\propto e^{iPn}\delta_{nm}$, we have recombined the diagonal parts in two Choy-Haldane states as in Eq.~(\ref{diagonal}) to make the Schr\"odinger equation satisfied, in which the diagonal term is proportional to the physical interaction, i.e. $UD_A$ or $UD_C$. This is a key part in obtaining the wave function as a sum of Choy-Haldane states. We will use this ``recombination'' technique below for the generic case.

\begin{figure}[h]
     \centering
     \begin{subfigure}[b]{0.8\linewidth}
        \centering
         \includegraphics[width=\linewidth]{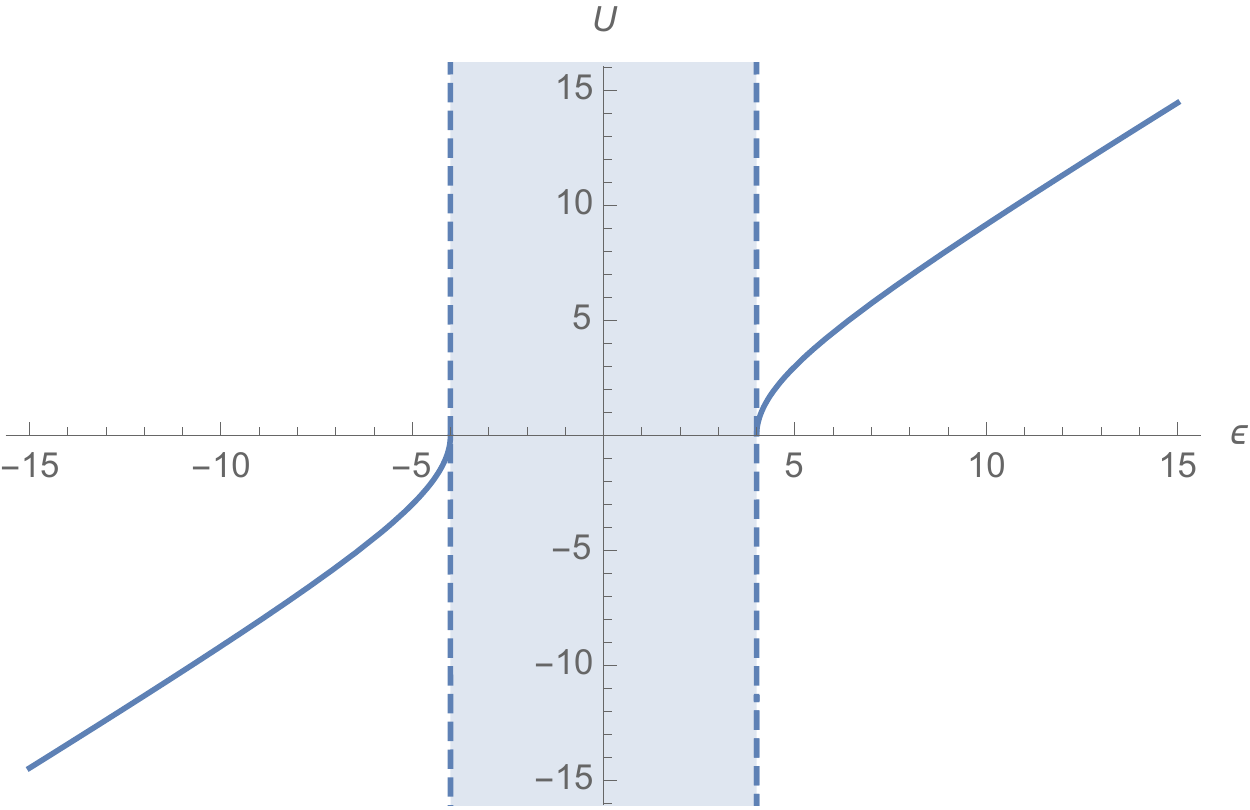}
         \caption{$A_{p,q}=-C_{p,q}$, $\ket{\psi_1}=\ket{HC_0}$; or $B=B_a$.}
         \label{type-1}
     \end{subfigure}
    \hfill
     \begin{subfigure}[b]{0.8\linewidth}
         \centering
         \includegraphics[width=\linewidth]{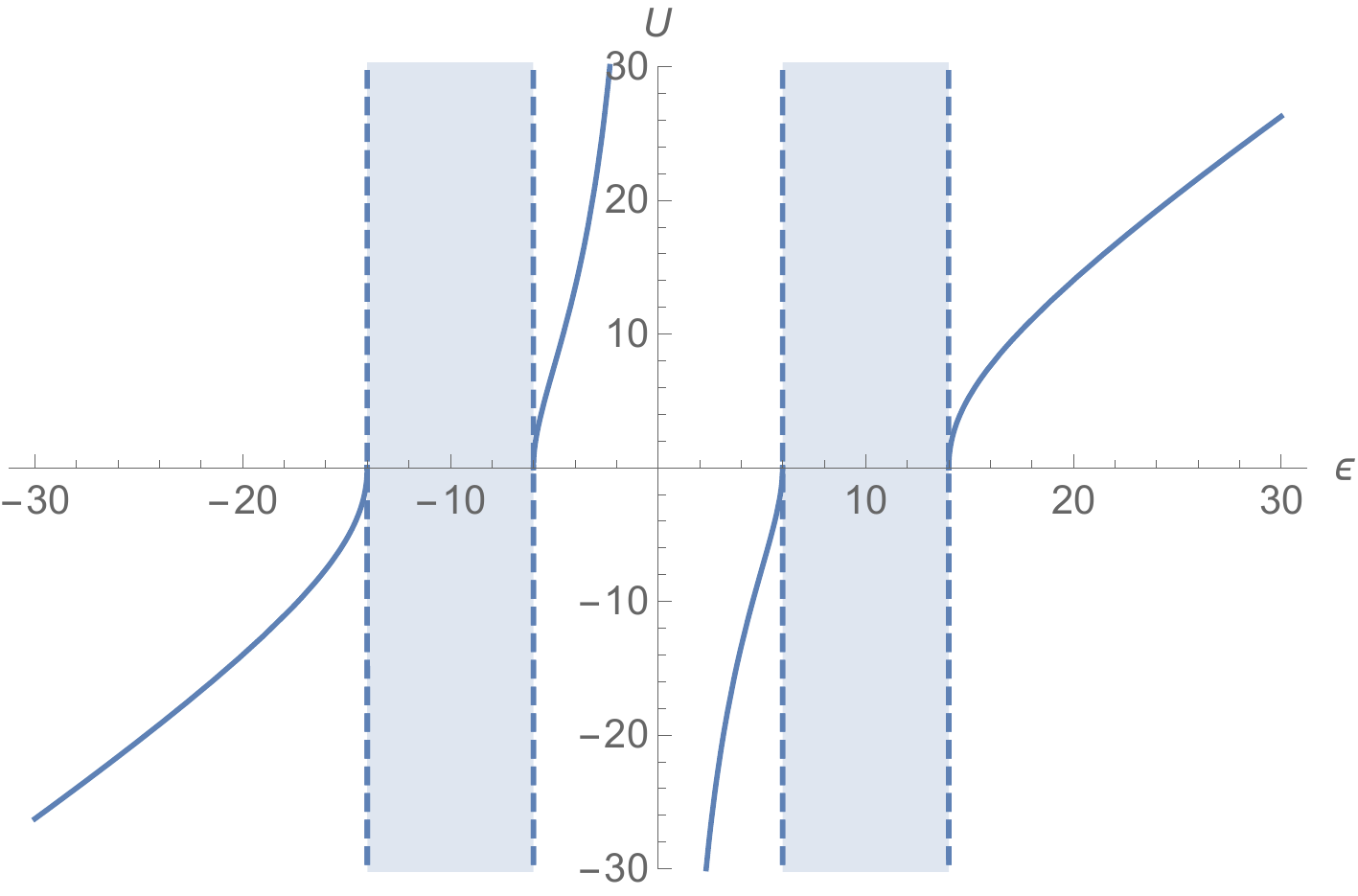}
         \caption{$A_{p,q}=C_{p,q}$, $\ket{\psi_2}=\ket{HC_1}+\lambda\ket{HC_2}$.}
         \label{cbh_spectrum_b}
     \end{subfigure}
     \caption{Spectrum of two copies of Bose-Hubbard model, two-excitation states in thermodynamic limit, $\Delta=0$, $J=1$, $\Omega=5$, $P=0$ for $U_1=U_2=U$.}
     \label{cbh_spectrum}
\end{figure}

\medskip \noindent {\bf Doublons}. In the thermodynamic limit ($N\rightarrow\infty$), we concentrate on those states outside of the continua, which are  called ``doublons''. Their wave functions are Choy-Haldane states with complex quasi-momenta: $k=a-iK$, $K>0$. For these states, we can directly write their energy equations given the total momentum $P$ and interaction strength $U$. There are three separate regions, by solving Eq.~(\ref{eqn:energy1}), Eq.~(\ref{eqn:energy2}), Eq.~(\ref{eqn:bethe}) and Eq.~(\ref{eqn:UU12}),
\begin{eqnarray}
    U&=-2\frac{\sqrt{+}\sqrt{-}}{\sqrt{+}+\sqrt{-}},\ & \mathrm{when} \ \epsilon<-2\Omega,\\
    U&=2\frac{\sqrt{+}\sqrt{-}}{\sqrt{+}-\sqrt{-}},\ & \mathrm{when} \ -2\Omega<\epsilon<2\Omega,\\
    U&=2\frac{\sqrt{+}\sqrt{-}}{\sqrt{+}+\sqrt{-}},\ & \mathrm{when} \ \epsilon>2\Omega.
    \label{doublon_dispersion}
\end{eqnarray}
where $\sqrt{\pm}\equiv\sqrt{(\epsilon\pm2\Omega)^2-16J^2\cos^2{\frac{P}{2}}}$. We plot the $\epsilon-U$ relation for $P=0$  in Fig.~\ref{cbh_spectrum_b}.

Notice that the doublon in between two continua lives in the continuum shown in Fig.~\ref{type-1} (In Sec.~\ref{sec:spectrum} we will name it as type-1 vacuum) as long as $U$ is large enough for any $\Omega$. This means that we have a bound state in the continuum.

\medskip
\noindent {\bf Antisymmetric solutions}. Up until now we have assumed that the antisymmetric part of $B$ vanishes, i.e., $B_a=0$. In fact there is a set of solutions with non-zero $B_a$. From the Schr\"odinger equations~(\ref{eqn:J1=J2}), if $A=C=0$, and $B=B_a$, the only remaining equation is 
\beq
\epsilon B_a + T B_a + B_a T =0. 
\eeq 
 
The solution of this equation is the antisymmetric Bethe state
\begin{equation}
   B_{n,m}= e^{i k n+i q m}-e^{i q n+i k m},
\end{equation}
with the Bethe constraint $e^{i k N}=1$. The corresponding energy is $\epsilon=\omega_k+\omega_q$. In the thermodynamic limit, these solutions form a new continuum in the spectrum, whose energy range coincides with the continuum formed by the case $A=-C$, as in Fig.~\ref{type-1}. However, the wave functions in the latter are orthogonal to those in the former case as their $B_a=0$.

\subsection{When $U_1\neq U_2$}
\label{sec:IVB}
We can now straightforwardly extend our result to the case when $U_1\neq U_2$. Notice from equation (\ref{scheoms}), $A_{p,q}$ and $C_{p,q}$ are essentially superpositions of $\ket{\psi_1}$ and $\ket{\psi_2}$. For instance, if $U_1=U$ and $U_2=0$, we have
\begin{eqnarray}
    A_p&=\frac{\epsilon-\omega_p-\omega_q-\eta_{pq}}{(\epsilon-\omega_p-\omega_q-\eta_{pq})(\epsilon-\omega_p-\omega_q-\eta_{pq})-\eta_{pq}^2} \frac{U A}{N},
    \\
    C_p&=\frac{\eta_{pq}}{(\epsilon-\omega_p-\omega_q-\eta_{pq})(\epsilon-\omega_p-\omega_q-\eta_{pq})-\eta_{pq}^2} \frac{U A}{N}.
\end{eqnarray}
where $\eta_{pq}=\frac{2\Omega^2}{\epsilon-\omega_p-\omega_q}$. We can write the above equations in another form,
\begin{eqnarray}
   \!\!\!\!\!\!\!\! A_p&=(\frac{1}{\epsilon-\omega_p-\omega_q-2\Omega}+\frac{1}{\epsilon-\omega_p-\omega_q+2\Omega}+\frac{1}{\epsilon-\omega_p-\omega_q}) \frac{U A}{2N},
    \\
  \!\!\!\!\! \!\!\! C_p&=(\frac{1}{\epsilon-\omega_p-\omega_q-2\Omega}+\frac{1}{\epsilon-\omega_p-\omega_q+2\Omega}-\frac{1}{\epsilon-\omega_p-\omega_q}) \frac{U A}{2N}.
\end{eqnarray}
From these, we can deduce that, in the position picture, we have
\begin{eqnarray}
A_{nm}&= \lambda'(HC_1+\lambda HC_2)+HC_0,
\\
C_{nm}&=\lambda'(HC_1+\lambda HC_2)-HC_0,
\end{eqnarray}
and 
\beq
B_{nm}=2\lambda'( HC_1-\lambda HC_2),
\eeq
where $\lambda$ and $\lambda'$ are relative weights to be determined.

The spectrum  should then possess three continua, outside of which there are three ``doublons''. According to the above wavefunction and the following Schr\"odinger equations they satisfy,
\begin{equation}
\label{eqn:U2=0}
    \left\{
    \begin{aligned}
        & \epsilon A + T A + A T -U D_A = \Omega B_s,
\\
& \epsilon B+ T B +B T=2\Omega (A+C),
\\
& \epsilon C + T C + C T = \Omega B_s,
    \end{aligned}
    \right.
\end{equation}
we have 
\begin{eqnarray}
\!\!\!\!\!\!&&\lambda'(\tilde{U}_1 s'_1+\lambda\tilde{U}_2 s'_2)+\tilde{U}_0 s'_0=U(\lambda'(s'_1+\lambda s'_2))+s'_0),
\\
\!\!\!\!\!\!&&\lambda'(\tilde{U}_1 s'_1+\lambda\tilde{U}_2 s'_2)-\tilde{U}_0 s'_0=0,
\\
\!\!\!\!\!\!&&\lambda'(\tilde{U}_1 s'_1-\lambda\tilde{U}_2 s'_2)=0,
\end{eqnarray}
where we have denoted $(1+s_i)$ by $s'_i$ for convenience. From the three equations, the interaction strength $U$ and the three other fictitious ones $\tilde{U}_i$'s can be related via $U={4}/({\frac{2}{\Tilde{U}_0}+\frac{1}{\Tilde{U}_1}+\frac{1}{\Tilde{U}_2}})$, similar to the previous example. 

This result is also confirmed by the numerical solution of the momentum-space Schr\"odinger equation; see the Eq.~(\ref{appeq:epsilonU}) for the relation between $\epsilon$ and $U$, which allows to numerically obtain $U$ given a $\epsilon$.

\begin{figure}[h]
\centering
\includegraphics[width=0.8\linewidth]{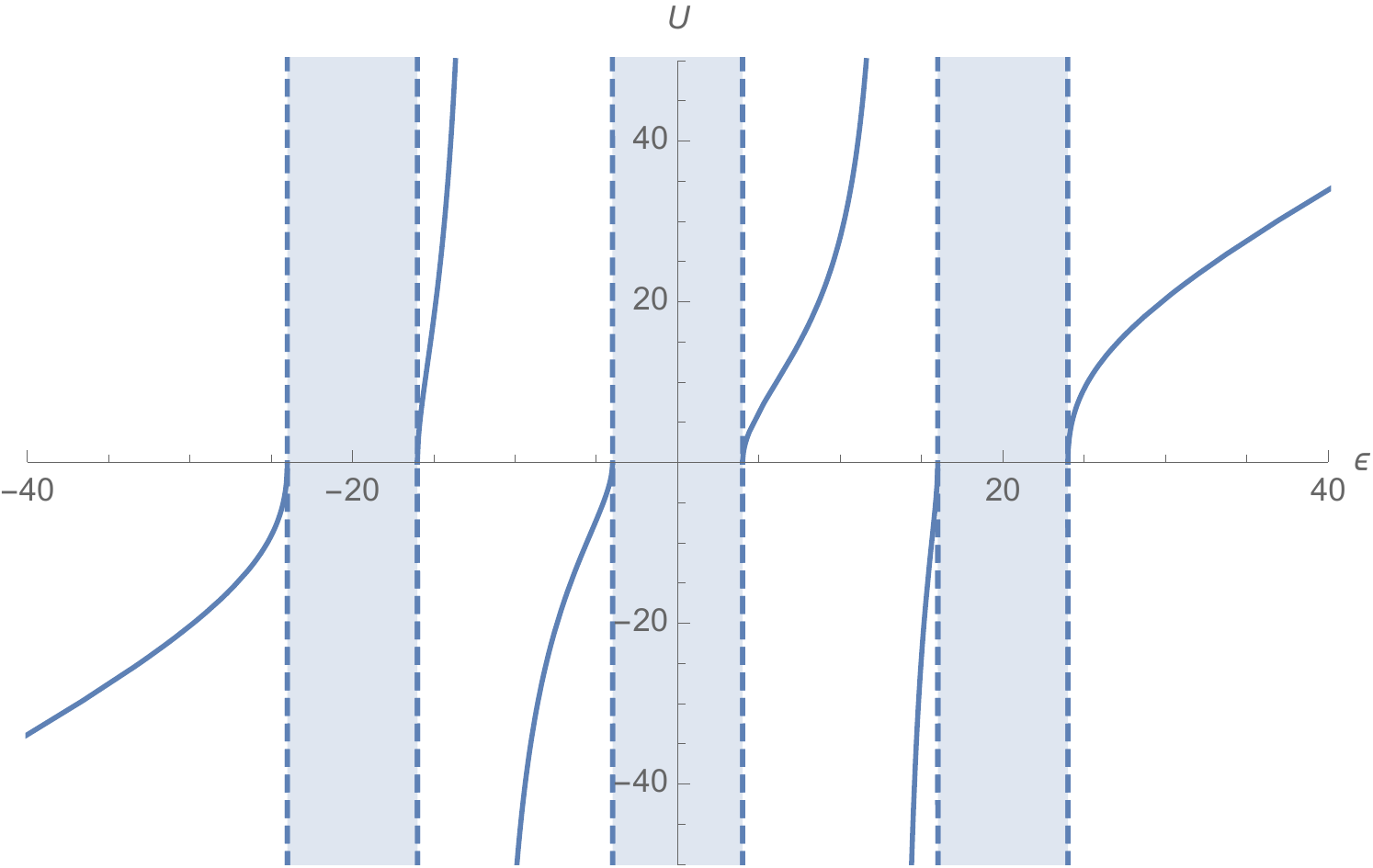}
\caption{When $U_2=0$ and $U_1=U$, $\Omega=10$, the spectrum possesses three continua, outside of which there are three ``doublons''.}
     \label{cbh_spectrum_c}
\end{figure}

Furthermore, we notice that the antisymmetric solutions ($B_a$) of the Schr\"odinger equations still remain when having arbitrary $U_1$ and $U_2$, because their $A$ and $C$ components are zero. Their wave functions and energies do not vary with any interaction. Then we can conclude that when two coupled Bose-Hubbard models with the same single-particle dispersion, there will be three types of two-excitation states in general. Two out of the three types are combinations of three different Choy-Haldane states, and the remaining contains  anti-symmetrized states with two atoms belonging to two different species. The ``doublon'' states are in the former two types.

\subsection{Generic case}
Using the insight from the simple, though already nontrivial, cases that we have just studied, we now move on to the generic case of the coupled BH model, anticipating  that the solutions are basically superposition of different Choy-Haldane states. We consider arbitrary intra-species interactions $U_1$ and $U_2$ and generally different hopping strengths $J_1$ and $J_2$ for the two respective species or copies of the BH model. (The inter-species interaction is not considered here in this section, but the solution to include it can be straightforwardly generalized; see Appendix~\ref{sec:VI}.) 

First we recall that the Schr\"odinger equations for this general case are
\begin{subequations}
\label{eq:generic}
\begin{eqnarray}
& \epsilon A + T_1 A + A T_1 -U_1 D_A = \frac{\Omega}{2} (B+B^T),
\\
& \epsilon B+ T_1 B +B T_2=2\Omega (A+C),
\\
& \epsilon C + T_2 C + C T_2 -U_2 D_C= \frac{\Omega}{2} (B+B^T).
\end{eqnarray}
\end{subequations}
We assume that there is a Choy-Haldane state $HC$ with quasi-momenta $k$ and $q$ is in $A$, denoted by $\lambda HC\subset A$, and that there is another corresponding term $\lambda' HC\subset C$ with the same $k$ and $q$. We would like to deduce how the  weights $\lambda$ and $\lambda'$ are related. We ignore the diagonal part for now, because the diagonal parts of different Choy-Haldane states could be recombined as done in Eq.~(\ref{diagonal}). Then from the equality of the first and third equations (via the $B$ part) above in Eq.~(\ref{eq:generic}), we can relate these two weights,
\beq
(\epsilon-\omega_k-\omega_q)\lambda=(\epsilon-\omega_k'-\omega_q')\lambda'.
\label{lambda}
\eeq
Since the matrix $B$ is not symmetric now, we can separate it into the symmetric and antisymmetric parts, 
\beq
B_s=\frac{1}{2}(B+B^T),~B_a=\frac{1}{2}(B-B^T).
\eeq
Therefore, there is also a corresponding Choy-Haldane state in $B_s$ as well: $\lambda\frac{\epsilon-\omega_k-\omega_q}{\Omega} HC\subset B_s$. 

As the right hand side of the second Schr\"odinger equation in the above Eq.~(\ref{eq:generic}) is manifestly symmetric as both $A$ and $C$ are, so should be the left hand side, and this gives,
\beq
&\epsilon B_s +T_1 B_s+B_s T_2 + \epsilon B_a +T_1 B_a+B_a T_2
\\
&= \epsilon B_s +T_2 B_s+B_s T_1 - \epsilon B_a - T_2 B_a - B_a T_1.
\eeq
We can simplify it to
\beq
&2\epsilon B_a + (T_1 + T_2) B_a + B_a (T_1 + T_2)
\\
&= (T_2 - T_1) B_s - B_s (T_2 - T_1)
\label{a-s},
\eeq
where $B_s$ contains the Choy-Haldane state $HC$: 
\begin{equation}
HC_{n,m}=\left\{
\begin{aligned}
&e^{i k n+i q m}+s_{k,q}e^{i q n+i k m}, \ & n\leq m,
\\
&HC_{m,n},  & n>m,
\end{aligned}
\right.
\end{equation}
where $s_{k,q}=\frac{2\big(\sin{k_i}-\sin{(P-k_i)}\big)-i\Tilde{u}_i}{2\big(\sin{k_i}-\sin{(P-k_i)}\big)+i\Tilde{u}_i}$ satisfies the corresponding Bethe equation. Substituting it into the RHS of Eq.~(\ref{a-s}),  we have, for $n<m$,
\beq
RHS_{n,m} = \chi (e^{i k n+i q m}-s_{k,q}e^{i q n+i k m}),
\eeq
where $\chi= \lambda\frac{\epsilon-\omega_k-\omega_q}{\Omega} (\omega_k-\omega_k'-\omega_q+\omega_q')$.

At this point, it seems  natural to introduce a corresponding term in $B_a$ of a similar form  in order to make the two sides equal, and  it is indeed the case with the following form, 
\begin{equation}
   HC'_{n,m}=\left\{
\begin{aligned}
& e^{i k n+i q m}-s_{k,q}e^{i q n+i k m}, \ & n< m
\\
& 0 , \ & n = m
\\
& -HC'_{m,n} , \ & n > m
\end{aligned}
\right.
\end{equation}
and  we denote the contribution of this Choy-Haldane state to $B_a$ as $\kappa \lambda\frac{\epsilon-\omega_k-\omega_q}{\Omega} HC' \subset B_a$. The weight $\kappa $ can be determined via equation~(\ref{a-s}),
\beq
\kappa=\frac{\omega_k-\omega_k'-\omega_q+\omega_q'}{2\epsilon-\omega_k-\omega_k'-\omega_q-\omega_q'}.
\label{eqn:kappa}
\eeq
Then from the second Schr\"odinger equation in Eq.~(\ref{eq:generic}) for $n<m-1$, we have
\begin{widetext}
\begin{eqnarray}
&LHS_{n,m}& = \lambda\frac{\epsilon-\omega_k-\omega_q}{\Omega}\big((2\epsilon-\omega_k-\omega_k'-\omega_q-\omega_q') - \kappa  (\omega_k-\omega_k'-\omega_q+\omega_q')\big)(e^{i k n+i q m}+s_{k,q}e^{i q n+i k m}),
\\
&RHS_{n,m}& = 4\Omega (\lambda+\lambda') (e^{i k n+i q m}+s_{k,q}e^{i q n+i k m}),
\end{eqnarray}
\end{widetext}
where $\lambda+\lambda'=\frac{2\epsilon-\omega_k-\omega_k'-\omega_q-\omega_q'}{\epsilon-\omega_k'-\omega_q'}\lambda$ according to equation~(\ref{lambda}). Equating both sides, we have the consistency equation that involves the energy, see Eq.~(\ref{eq:1ex}),
%\beq
\begin{eqnarray}
& & (2\epsilon-\omega_k-\omega_k'-\omega_q-\omega_q')^2 -  (\omega_k-\omega_k'-\omega_q+\omega_q')^2 \nonumber\\
&=&4\Omega^2\frac{(2\epsilon-\omega_k-\omega_k'-\omega_q-\omega_q')^2}{(\epsilon-\omega_k-\omega_q)(\epsilon-\omega_k'-\omega_q')}.
%\eeq
\end{eqnarray}
The solutions (four of them in general) of this energy equation are exactly a sum of two single-excitation state energies,
\beq
\epsilon=\epsilon_k^{\pm}+\epsilon_q^{\pm},
\label{energyeq}
\eeq
where $\epsilon_k^{\pm}=\big(\omega_k+\omega'_k\pm\sqrt{(\omega_k-\omega'_k)^2+4\Omega^2}\big)/2$. But we emphasize that  $k$ and $q$ are quasi-momenta instead of physical momenta and their sum is the total momentum $P$, i.e., $k+q=P$. We remark that when $J_1=J_2$ and hence $\omega_k=\omega'_k$ (if $\Delta=0$) , Eq.~(\ref{energyeq}), with $U_1=U_2$, reduces to (a) Eqs.~(\ref{eqn:energy1}) and~(\ref{eqn:energy2}) for the case of $A_{p,q}=C_{p,q}$, as well as (b) $\epsilon=\omega_{k}+\omega_{P-k}$ for the case of $A_{p,q}=-C_{p,q}$ in Sec.~\ref{sec:IVA}.

Indeed, after specifying the total momentum $P$, we can have \textit{four} sets of quasi-momenta $(k_i,q_i=P-k_i)$ in total labeled by the index $i$  for some  energy $\epsilon$ satisfying Eq.~(\ref{energyeq}).  The wave functions of two-excitation eigenstates are combinations of the four corresponding Choy-Haldane states,
\begin{equation}
    \left\{
    \begin{aligned}
       A &= \sum_{i=1}^4 \lambda_i HC_i,
\\
B &= \sum_{i=1}^4 \lambda_i \frac{\epsilon-\omega_{k_i}-\omega_{q_i}}{\Omega} (HC_i + \kappa_i HC_i' ),
\\
C &= \sum_{i=1}^4 \lambda_i' HC_i.
    \end{aligned}
    \right.
    \label{eqn:choy-haldane}
\end{equation}
There are four equations determining weights $\lambda_i$,
\begin{eqnarray}
&& \sum_{i=1}^4 J_1 \Tilde{u}_i \lambda_i (1+s_i)=U_1 \sum_{i=1}^4 \lambda_i (1+s_i),
\\
&& \sum_{i=1}^4 J_2 \Tilde{u}_i \lambda'_i (1+s_i)=U_2 \sum_{i=1}^4 \lambda'_i (1+s_i),
\\
&& \sum_{i=1}^4 \kappa_i \lambda_i (\epsilon-\omega_{k_i}-\omega_{q_i}) (1-s_i)=0,
\\
%&\begin{aligned}
&& \sum_{i=1}^4 \lambda_i (\epsilon-\omega_{k_i}-\omega_{q_i})\big((J_1+J_2) \Tilde{u}_i  (1+s_i)  \nonumber\\
 &&\quad  - (J_1-J_2)2\mathrm{i}(\sin{k_i}+\sin{q_i})  \kappa_i (1-s_i)  \big)=0.
%\end{aligned}
\end{eqnarray}
The virtual interaction strength $\tilde{u}_i$ (for $i=1,2,3,4$) depends on the quasi-momentum $k_i$ and the total momentum $P$ via the Bethe equations. While the non-diagonal entries of the matrix Schr\"odinger equations   give us energy equations Eq.~(\ref{energyeq}), as we have seen above, these four equations regarding $\lambda$'s come from the almost diagonal entries ($n=m$ and $n=m-1$) of the Schr\"odinger equations.

To recapitulate, we show that from a specific energy $\epsilon$, we obtain four sets of quasi-momenta $\{k_i, q_i\}$. The total two-particle wave function is composed of four Choy-Haldane states, each having a set of quasi-momenta $\{k_i, q_i\}$ and each with a weight $\lambda_i$ determined by the above equations. In the meantime, we see 4 continua in the two-particle spectrum (2 of the 4 continua coincide). There is, however, one exception in this description: when $P=0$, instead of four sets of solutions, we only have three sets of solutions from Eq.~(\ref{energyeq}). This is due to the fact that the two quasi momenta are opposite: $q=P-k=-k$ and thus the four equations in Eq.~(\ref{energyeq}) are not all independent:
\beq
\epsilon^{+}_k+\epsilon^-_{-k}\equiv\epsilon^{-}_k+\epsilon^+_{-k}.
\eeq

When $P=0$, it turns out that  besides the states in the form $\sum_{i} HC_i$, there are also anti-symmetric solutions of $B$ (with $A=C=0$) as we saw earlier when $J_1=J_2$. We now check this statement.  The reduced equations for $B$ by taking $A=C=0$ become 
\begin{eqnarray}
2\epsilon B_a + (T_1+T_2) B_a + B_a (T_1+T_2) & = 0,
\\
 (T_1-T_2) B_a - B_a (T_1-T_2) & = 0.
\end{eqnarray}
When $T_1\neq T_2$, these two equations can be replaced by (as $T_1=(J_1/J_2)T_2$)
\begin{eqnarray}
(T_1+T_2)B_a=-\epsilon B_a,
\\
B_a(T_1+T_2)=-\epsilon B_a.
\end{eqnarray}
Matrix $T_1+T_2$ has eigenvalues $\lambda=-2(J_1+J_2)\cos{k}$, where $k=\frac{2r\pi}{N}$, with $r=1,\dots,N$. So when $k\neq0,\pi$, there is degeneracy: $(e^{ik},\dots,e^{iNk})$ and $(e^{-i k},\dots,e^{-iNk})$ are distinct solutions but have the same eigenvalue w.r.t. $T_1+T_2$. Therefore, we can take the antisymmetric direct product of the two vectors to form a solution of the above matrix equations
\beq
(B_a)_{i,j}=\sin\big((i-j)k\big).
\label{eqn:anti}
\eeq
This solution corresponds to two particles where one has momentum $k$ and the other $-k$, and thus we have a continuum formed by anti-symmetric states when total momentum $P=k+(-k)=0$. 

The existence of this continuum can also be understood as follows. When $P=0$, from Eq.~(\ref{eqn:kappa}), we have $\kappa=0$, as $\omega_k=\omega_{-k}$, and that Eq.~(\ref{a-s}) reduces to $(T_1-T_2)B_s=B_s(T_1-T_2)$. Thus the wave function components in $B$ of Eq.~(\ref{eqn:choy-haldane}) will have a vanishing anti-symmetric part. As we obtain 3 independent solutions from Eq.~(\ref{energyeq}), the number of continua in the form of combinations of Choy-Haldane states is reduced from 4 to 3. Meanwhile, we have anti-symmetric states shown in Eq.~(\ref{eqn:anti}). In total we still have $3+1=4$ sets of continua when $N\rightarrow\infty$. 

When $P\neq0$, there are 4 continua formed by combinations of Choy-Haldane states. So  when $P\neq0$, the anti-symmetric $B_a$ itself is no longer the eigenstate of the system. Instead, as $\kappa\ne0$, it becomes an component in the Choy-Haldane form of wave functions as in Eq.~(\ref{eqn:choy-haldane}), which makes the number of such continua be 4. Note that what happens in the $J_1=J_2$ case is that, $\kappa\equiv0$ for all $P$ because $\omega_k\equiv\omega'_k$. Thus the $B_a$ continuum independently exists for all $P$, as we saw earlier in Sec.~\ref{sec:IVA}.

\begin{figure}[h]
     \centering
     \begin{subfigure}[b]{0.8\linewidth}
       \centering
\includegraphics[width=0.9\linewidth]{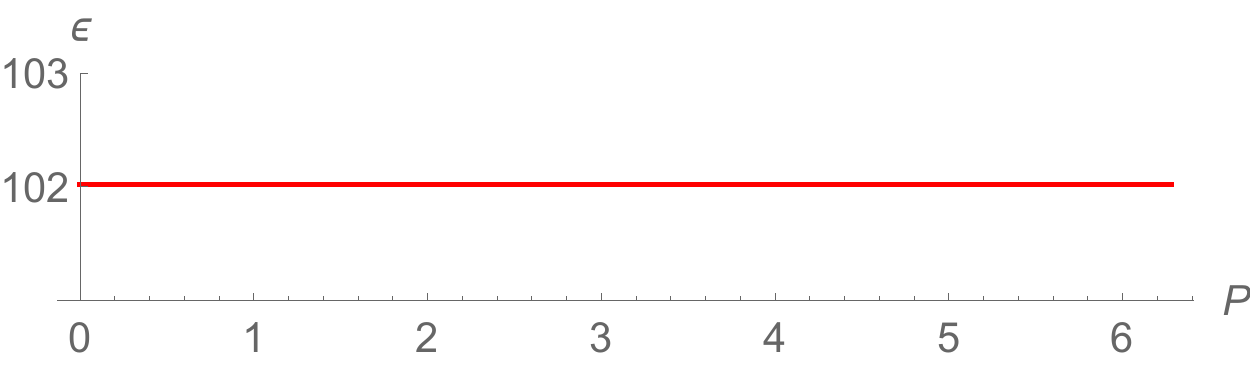}
\caption*{}
     \end{subfigure}
    \hfill
     \begin{subfigure}[b]{0.8\linewidth}
         \centering
\includegraphics[width=0.9\linewidth]{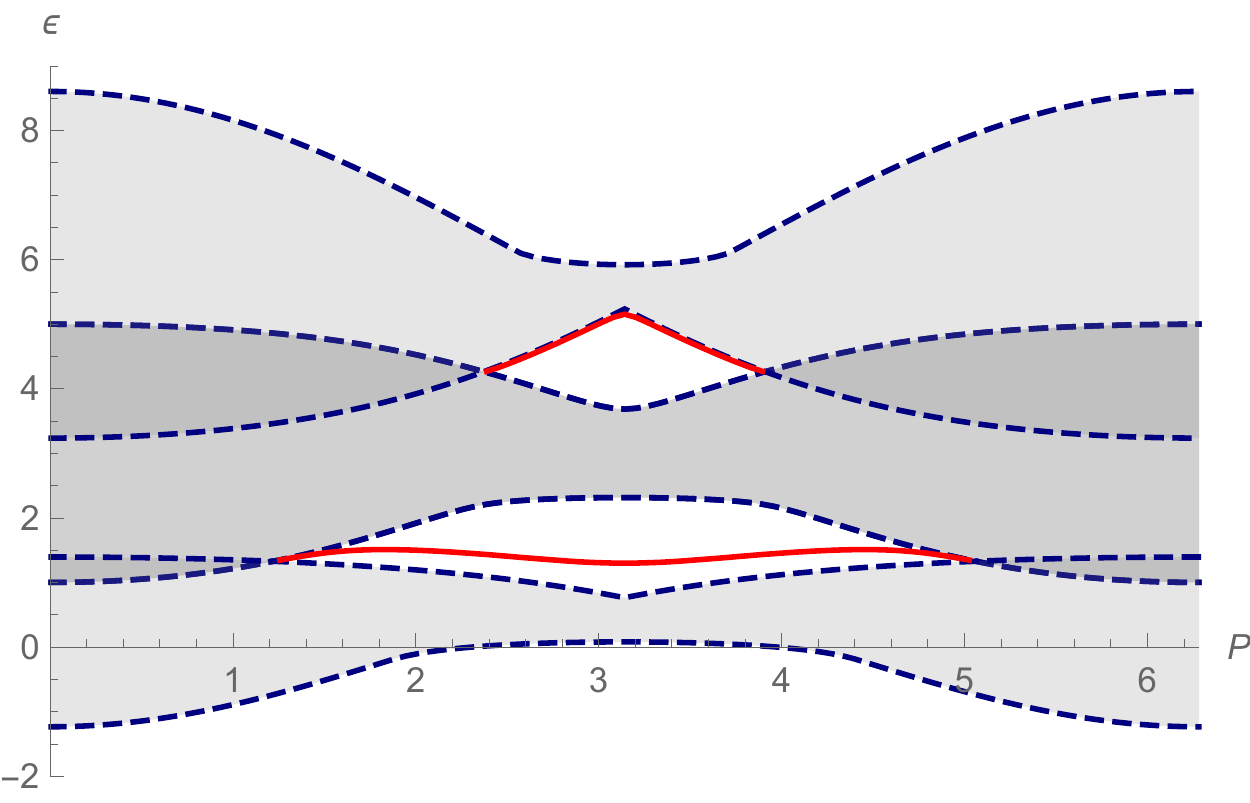}
\caption*{}
     \end{subfigure}
     \caption{The spectrum of the 2-particle states for $J_1=0, U_1=100J_2$, $U_2=0$, and $\Delta=\Omega=J_2$. The red solid curves represent the dispersion for doublons. The third doublon is around $\epsilon\approx102.2J$ (shown on top), and its momentum dependence is extremely weak.}
     \label{delta_spectrum}
\end{figure}

\smallskip \noindent {\bf An example}. We now give an explicit spectrum of the system when $J_1=0,  U_1=100J_2$, $\Omega=J_2$, $\Delta=J_2$ and $U_2=0$ in Fig.~\ref{delta_spectrum}. Notice that the middle continuum has a darker shade than the other two, as it represents (effectively) two continua. These two of the four continua coincide, while the doublons appear in between different continua, plotted in solid red. The third doublon has a very high energy ($\epsilon\approx100J_2$) and its dispersion is almost flat.  We note that similar plots were also shown in Ref.~\cite{Shi2018Oct}, in which the authors used perturbations to acquire the two-particle spectrum. 

In this case, since $J_1=0$, we have 3 sets of solutions and 3 corresponding Choy-Haldane states to combine. But different from the $P=0$ case, where we also have 3 sets of solutions, the number of continua in the spectrum is still 4. The reason is that, in the $P=0$ case, two solutions of the energy equations coincide, while when $J_1$ approaches $0$, one of the four solutions moves to $\infty$. When the quasi-momentum $k$ goes to $\infty$, the Choy-Haldane state ``does not'' disappear. But its weight $\lambda$ for every eigenstate goes to $0$. Therefore the number of continua does not change and remains 4.

\smallskip\noindent {\bf Including the inter-species interaction}. We can also generalize our method discussed in this section to include the inter-species interaction. The steps are similar and the results are presented in Appendix~\ref{sec:VI}.

\section{Doublon states}
\label{sec:doublon}
As mentioned earlier already, we call the states outside of the continua doublon states. In the single-species Bose-Hubbard model, there is one state outside of the continuum. From Eq.~(\ref{betheeq}), in the thermodynamic limit, $U=\pm4J\cos{\frac{P}{2}}\sinh{K}$, and from the energy equation, $\epsilon=\pm4J\cos{\frac{P}{2}}\cosh{K}$. In the limit of $U\rightarrow\infty$, this state represents two atoms residing on the same site, and traveling through the lattice with momentum $P$. Thus, this state is called a doublon and  can be regarded as one particle with dispersion relation $\epsilon=\pm\sqrt{U^2+16J^2\cos^2{\frac{P}{2}}}$.

With our analytic results in the previous section, we can easily study doublons in  the case of the coupled Bose-Hubbard model. When $N\rightarrow\infty$, the doublon states have the following real-space wave functions, e.g. the   components of $A$,
\beq
A_{n,m}=\sum_{i=1}^4 \lambda_i\exp{(-K_i|n-m|)}.
\eeq

\begin{figure}[h]
     \centering
     \begin{subfigure}[b]{0.75\linewidth}
        \centering
     \includegraphics[width=\linewidth]{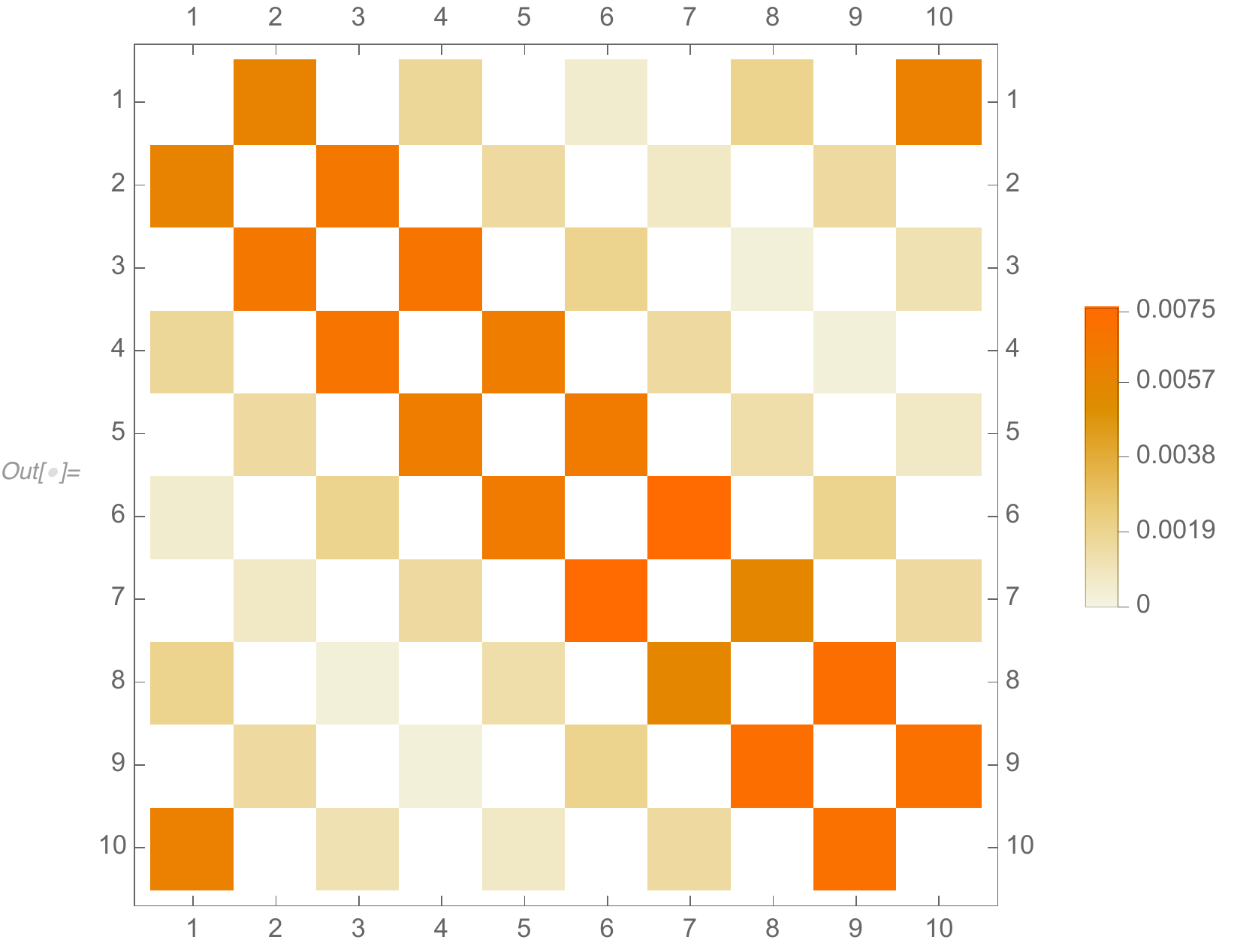}
         \caption{ $|A_{ij}|^2$ (or $|C_{ij}|^2$),}
     \end{subfigure}
    \hfill
     \begin{subfigure}[b]{0.8\linewidth}
         \centering
     \includegraphics[width=\linewidth]{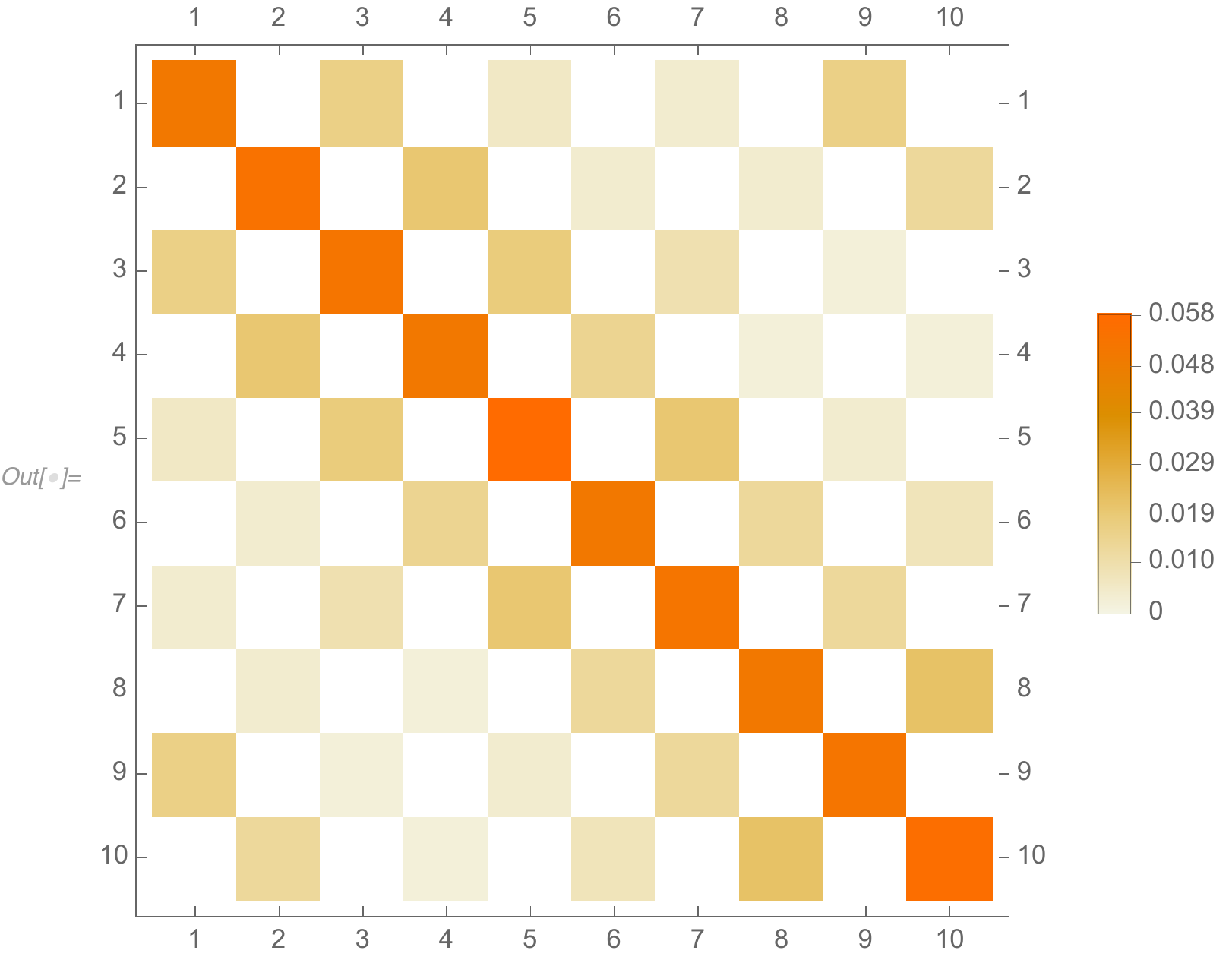}
     \caption{ $|B_{ij}|^2$ .}
     \end{subfigure}
     \caption{Matrix plot of a finite-energy doublon state when $U_1=U_2\rightarrow\infty$ and $\Omega/J=2$, in which case when the diagonal entries of $A$ and $C$ are $0$. We take $N=10$ for the plot. }
     \label{fig:doublon}
\end{figure}

As an illustration, we focus on the properties of doublons in the limit of infinite  interactions (i.e., $U_1$ and $U_2$ being very large). In the general case ($J_1\neq J_2$), from the Schr\"odinger equations we have
\begin{equation}
    \left\{
    \begin{aligned}
       &\sum_{i=1}^4 J_1 \Tilde{u}_i \lambda_i =U_1 \sum_{i=1}^4 \lambda_i, 
\\
&\sum_{i=1}^4 J_2 \Tilde{u}_i \lambda'_i =U_2 \sum_{i=1}^4 \lambda'_i,
\\
& \sum_{i=1}^4 \lambda_i (\epsilon-\omega_{k_i}-\omega_{q_i})\big((J_1+J_2) \Tilde{u}_i 
\\
&  \quad - (J_1-J_2)2\mathrm{i}(\sin{k_i}+\sin{q_i})  \kappa_i  \big)=0,
\\
&\sum_{i=1}^4 \kappa_i \lambda_i (\epsilon-\omega_{k_i}-\omega_{q_i}) =0.
    \end{aligned}
    \right.
\end{equation}
When $U_1\rightarrow\infty$, the r.h.s. of the first equation must vanish. Therefore, $\sum_i \lambda_i\rightarrow0$. Two of the three doublon dispersions have finite energies in this limit, their diagonal parts of $A$ in the wave function vanish, while diagonal parts of $B$ and $C$  are the dominant components of the wave functions. When $U_2\rightarrow\infty$ as well, the diagonal parts of $A$ and $C$ will both vanish for the one of the three doublon dispersions that still has finite energy, while the diagonal parts of $B$ dominate the wave functions, 
\beq
B_{n,m} = \sum_{i=1}^4 \lambda''_i \big(1 + \kappa_i \mathrm{sign}(n-m)\big)\exp{(-K_i|n-m|)},
\eeq
where $\lambda''_i\equiv \lambda_i \frac{\epsilon-\omega_{k_i}-\omega_{q_i}}{\Omega}$. We show in Fig.~\ref{fig:doublon} the probability density from the wave function 
for $U_1=U_2\rightarrow\infty$, which clearly demonstrate the adjacency feature.

We have seen in the last section that the number of solutions of energy equations can be deduced by the number of independent linear equations on $\lambda_i$. The number of continua can be further inferred. While number of doublons is fixed to be 3, their asymptotic energy as interactions become large can be simply derived. When $U_1$ and $U_2$ are large, in the leading order we can ignore the hopping strength $J_i$. Then the matrix equations are reduced to scalar equations,
\beq
\begin{pmatrix}
\epsilon-U_1 & -\Omega &
\\
-2\Omega & \epsilon & -2\Omega
\\
& -\Omega & \epsilon-U_2
\end{pmatrix}
\begin{pmatrix}
a
\\
b
\\
c
\end{pmatrix}
\approx0.
\eeq

If $U_1$ and $U_2$ are both large, two doublons will have energy $U_1$, $U_2$ and the other one has a finite energy. If $U_2=0$ and $U_1$ large, one doublon will have energy $U_1$ and the other two have a finite energy. We have seen these two cases in Sec.~\ref{sec:mainresults}. We can also study the leading order energy of the three doublons when including inter-species interaction and $\Omega\gg J$ in this way.

\section{Detailed analysis of discrete spectrum}
\label{sec:spectrum}
In this section, we will describe all two-excitation  eigenstates and examine their properties, such as the inverse participation ratio and entanglement behavior. To do these in detail, we will focus on certain limits of the interaction.

When $U_{1}=U_{2}=U\rightarrow\infty$, we have three types of solutions: 
\begin{enumerate}
    \item 
 $\psi_{1}$: $A=-C$ is a Choy-Haldane state with $s_{k,-k}=-1$, and $B=0$, $\epsilon=2\omega_{k}$; 
\item
 $\psi_{2}$: $A=C=HC_{1}+\lambda HC_{2}$, and $B=2(HC_{1}-\lambda HC_{2})$, with the energy equation being $\epsilon=2\omega_{k_{1}}+2\Omega=2\omega_{k_2}-2\Omega$; 
\item
$\psi_3$: anti-symmetrized states, which will not be of our concern.
\end{enumerate}

The first type is a single Choy-Haldane state. When $U\rightarrow\infty$, from the Bethe equation $e^{-ikN}=-1$, the quasi-momentum $k=\frac{2\pi}{N}\times(\mathrm{Half\,integer})$.

\begin{figure}[h]
\centering
\includegraphics[width=0.8\linewidth]{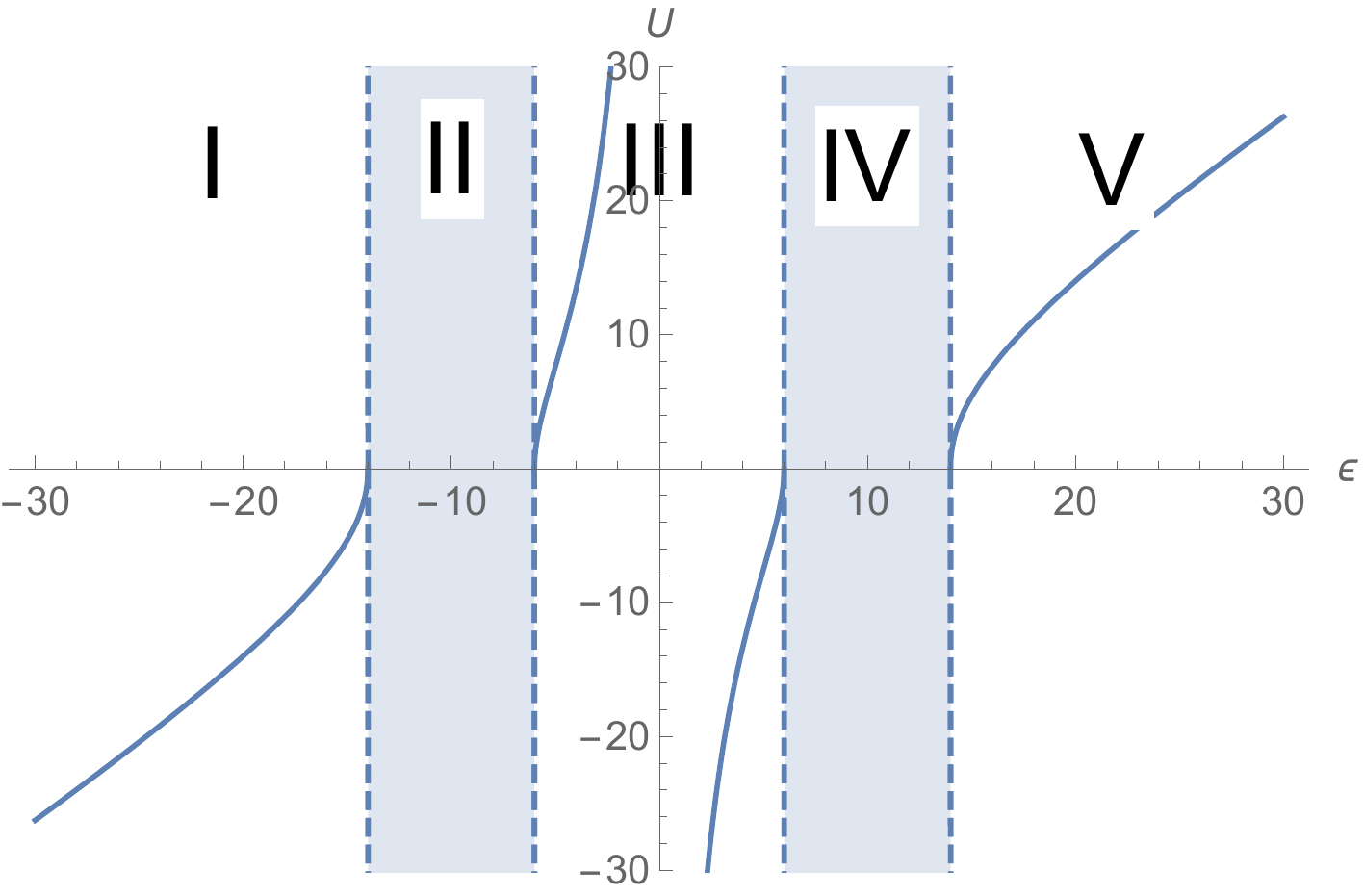}
\caption{The spectrum of the second type of states can be divided into 5 parts. This is the spectrum for $P=0$ specifically.}
\label{cbh_spect}
\end{figure}

The second type of states contain two Choy-Haldane states. It is shown in the previous section that the spectrum of this type of states possess two continua and two doublon dispersions. We divide the spectrum into five parts as in Fig.~\ref{cbh_spect} and discuss them one by one. We note that the spectrum is different when $\Omega/J<2$, in which case the two continua overlap. In this case, the only qualitative difference is in region III: instead of having a localized doublon state, there are additional extended states in the overlapping continuum in region III.

We list the eigen-energies and the wave functions below.
\begin{enumerate}
    \item Region I:
    \beq
    \epsilon&=-4\cos{\frac{P}{2}}\cosh{K_1}+2\Omega
    \\
    &=-4\cos{\frac{P}{2}}\cosh{K_2}-2\Omega
    \eeq
    and
    \beq
    A&=C=HC_1+\lambda HC_2,
    \\
    B&=2(HC_1-\lambda HC_2),
    \eeq
    where $\lambda=-\frac{1+s_1}{1+s_2}=-\frac{1+e^{-i(\frac{P}{2}-iK_1) N}}{1+e^{-i(\frac{P}{2}-iK_2) N}}$. When $U\rightarrow-\infty$, $K_1$ and $K_2\rightarrow\infty$; when $U\rightarrow+\infty$, there is no solution. 
    \item Region II:
    \beq
    \epsilon&=-4\cos{\frac{P}{2}}\cosh{K}+2\Omega
    \\&=-2\cos{k}-2\cos{(P-k)}-2\Omega
    \eeq
    and
    \beq
    A&=C=HC_1+\lambda HC_2,
    \\
    B&=2(HC_1-\lambda HC_2),
    \eeq
    where $\lambda=-\frac{1+s_1}{1+s_2}=-\frac{1+e^{-ik N}}{1+e^{-i(\frac{P}{2}-iK) N}}$. When $U\rightarrow+\infty$, $\tilde{u}_1=-\tilde{u}_2$, whose expressions are given as follows,
    \beq
  \label{eq:u1u2}
  \tilde{u}_1&=2(\sin{k}-\sin{(P-k)})\tan{\frac{kN}{2}},
    \\
    \tilde{u}_2&=-4i\cos{\frac{P}{2}}\sinh{K}\tan{(\frac{P}{2}-iK)\frac{N}{2}}.
    \eeq
    Given a total site number $N$, the total momentum can be $P=\frac{2\pi r}{N}$, with $r=0,1,\dots,N-1$. For each $P$ value, the above equations could be solved numerically. For instance, we take $N=10$ and $\Omega=10J$. When $P=0$, we plot in Fig.~\ref{region2_0} the l.h.s. and r.h.s. of the equation $\tilde{u}_1=-\tilde{u}_2$ in terms of $k$.
\begin{figure}[h]
     \centering
     \begin{subfigure}[b]{0.8\linewidth}
        \centering
         \includegraphics[width=\linewidth]{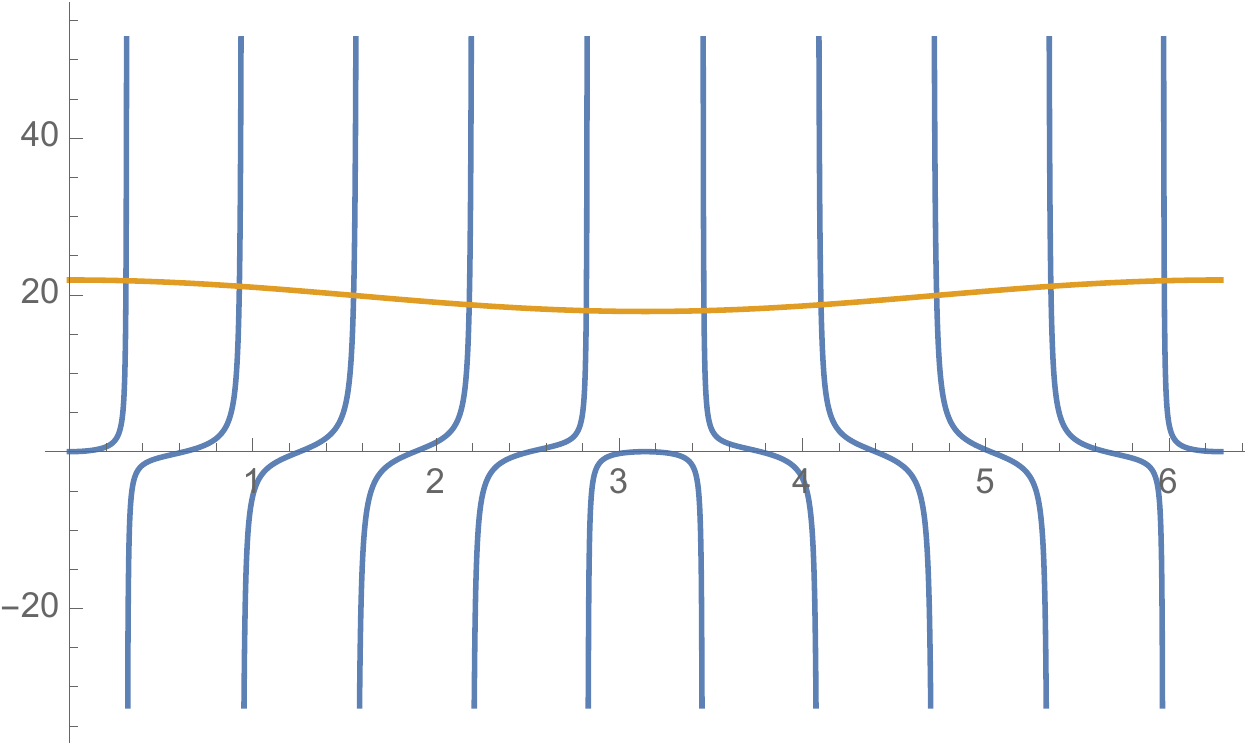}
         \caption{$P=0$,}
         \label{region2_0}
     \end{subfigure}
    \hfill
     \begin{subfigure}[b]{0.8\linewidth}
         \centering
         \includegraphics[width=\linewidth]{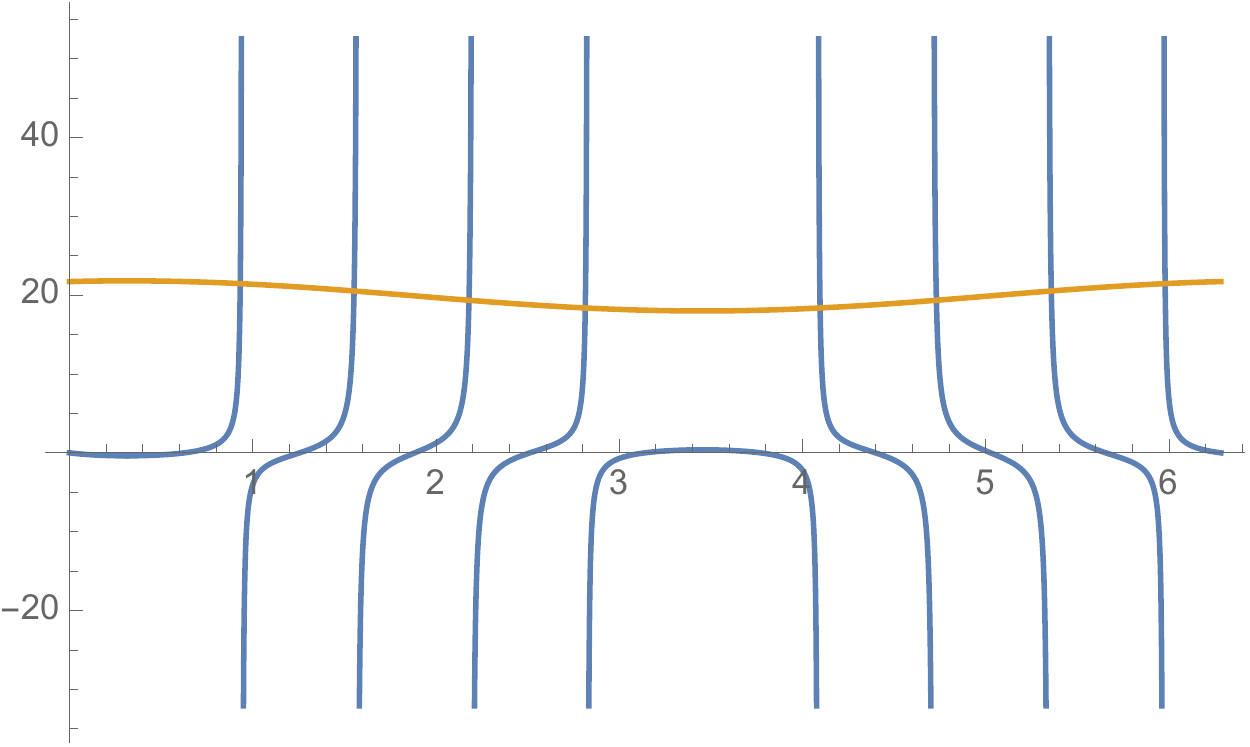}
         \caption{$P=\frac{2\pi}{10}$.}
         \label{region2_1}
     \end{subfigure}
     \caption{\label{fig:region2}Solving $\tilde{u}_1=-\tilde{u}_2$ in Eq.~(\ref{eq:u1u2}). The blue (yellow) curve is the l.h.s. (r.h.s.) of the equation 
%$\tilde{u}_1=-\tilde{u}_2$ 
in terms of $k$, for $N=10$ type-2 states in region II. Every intersection of the two curves gives one such state.}
\end{figure}
    Thus the equations have 5 solutions, if we consider the fact that exchanging $k$ and $P-k$, we essentially have the same state. When $P=\frac{2\pi}{10}$, the corresponding plot is shown in Fig.~\ref{region2_1}. There are 4 solutions in this case. This pattern persists if we continue with higher $P$. When $r$ is even, we would have distinct 5 states, but when $r$ is odd, we would have only 4 distinct solutions. In total, when $N=10$, region II of the second type gives 55 states. 
    \item Region III:
    \beq
    \epsilon&=-4\cos{\frac{P}{2}}\cosh{K_1}+2\Omega
    \\
    &=4\cos{\frac{P}{2}}\cosh{K_2}-2\Omega
    \eeq
    and
    \beq
    A&=C=HC_1+\lambda HC_2,\\
    B&=2(HC_1-\lambda HC_2),
    \eeq
    where $\lambda=-\frac{1+s_1}{1+s_2}=-\frac{1+e^{-i(\frac{P}{2}-iK_1) N}}{1+e^{-i(\pi+\frac{P}{2}-iK_2) N}}$. When $U\rightarrow+\infty$, $\epsilon=0$, the equation $\tilde{u}_1=-\tilde{u}_2$ is satisfied automatically. The energy equations become $2\cos{\frac{P}{2}}\cosh{K}=\Omega$ where$K_1=K_2=K$.
    \begin{eqnarray}
    \tilde{u}_1&=&-4i\cos{\frac{P}{2}}\sinh{K}\tan{\big((\frac{P}{2}-iK)\frac{N}{2}\big)}
    \\
    \tilde{u}_2&=&4i\cos{\frac{P}{2}}\sinh{K}\tan{\big((\pi+\frac{P}{2}-iK)\frac{N}{2}\big)}
    \end{eqnarray}
    We notice that the factor $\tan{\big((\frac{P}{2}-iK)\frac{N}{2}\big)}=\tan{\big((\pi+\frac{P}{2}-iK)\frac{N}{2}\big)}$ for any $N$, and therefore $\tilde{u}_1=-\tilde{u}_2$. There is 1 state for each $P$, thus 10 states in total. 
    
    If the two continua II and IV overlap, as long as $\Omega\neq0$, the above equations will still give us some valid wave functions. However, in this case we would expect some extended states in the overlapped region. The quasi-momenta of which are real.
    \beq
    \epsilon&=-2\cos{k_1}-2\cos{(P-k_1)}-2\Omega
    \\
    &=-2\cos{k_2}-2\cos{(P-k_2)}+2\Omega
    \eeq
    while
    \beq
    \tilde{u}_1&=2(\sin{k_1}-\sin{(P-k_1)})\tan{\frac{k_1N}{2}}.
     \\
    \tilde{u}_2&=2(\sin{k_2}-\sin{(P-k_2)})\tan{\frac{k_2N}{2}}.
    \eeq

    \item Region IV:
     \beq
    \epsilon&=-2\cos{k}-2\cos{(P-k)}+2\Omega
    \\
    &=4\cos{\frac{P}{2}}\cosh{K}-2\Omega
    \eeq
    and
    \beq
    A&=C=HC_1+\lambda HC_2,\\
    B&=2(HC_1-\lambda HC_2),
    \eeq
    where $\lambda=-\frac{1+s_1}{1+s_2}=-\frac{1+e^{-i(\pi+\frac{P}{2}-iK) N}}{1+e^{-ik N}}$. When $U_1\rightarrow+\infty$, $\tilde{u}_1=-\tilde{u}_2$, where
    \beq
    \tilde{u}_1&=-4i\cos{\frac{P}{2}}\sinh{K}\tan{(\frac{P}{2}-iK)\frac{N}{2}}
     \\
    \tilde{u}_2&=2(\sin{k}-\sin{(P-k)})\tan{\frac{kN}{2}}.
    \eeq
    As in region II, there are 55 states in total when $N=10$.
    \item Region V:
     \beq
    \epsilon&=4\cos{\frac{P}{2}}\cosh{K_1}+2\Omega
    \\
    &=4\cos{\frac{P}{2}}\cosh{K_2}-2\Omega
    \eeq
    and
    \beq
    A&=C=HC_1+\lambda HC_2,
    \\
    B&=2(HC_1-\lambda HC_2),
    \eeq
    where $\lambda=-\frac{1+s_1}{1+s_2}=-\frac{1+e^{-i(\pi+\frac{P}{2}-iK_1) N}}{1+e^{-i(\pi+\frac{P}{2}-iK_2) N}}$.
    When $U_1\rightarrow+\infty$, $K_1$ and $K_2\rightarrow\infty$. 
\end{enumerate}
In region III and V, eigen-wave functions are straightforwardly obtained. In region II and IV, we need to solve for the two quasi-momenta satisfying both energy equations and the equation $\tilde{u}_1=-\tilde{u}_2$, as illustrated in Fig.~\ref{fig:region2}.

\smallskip
Now with solutions of all two-particle eigenstates, we examine them in terms of their inverse participation ratio.

\begin{figure}[h]
     \centering
     \begin{subfigure}[b]{0.8\linewidth}
        \centering
         \includegraphics[width=\linewidth]{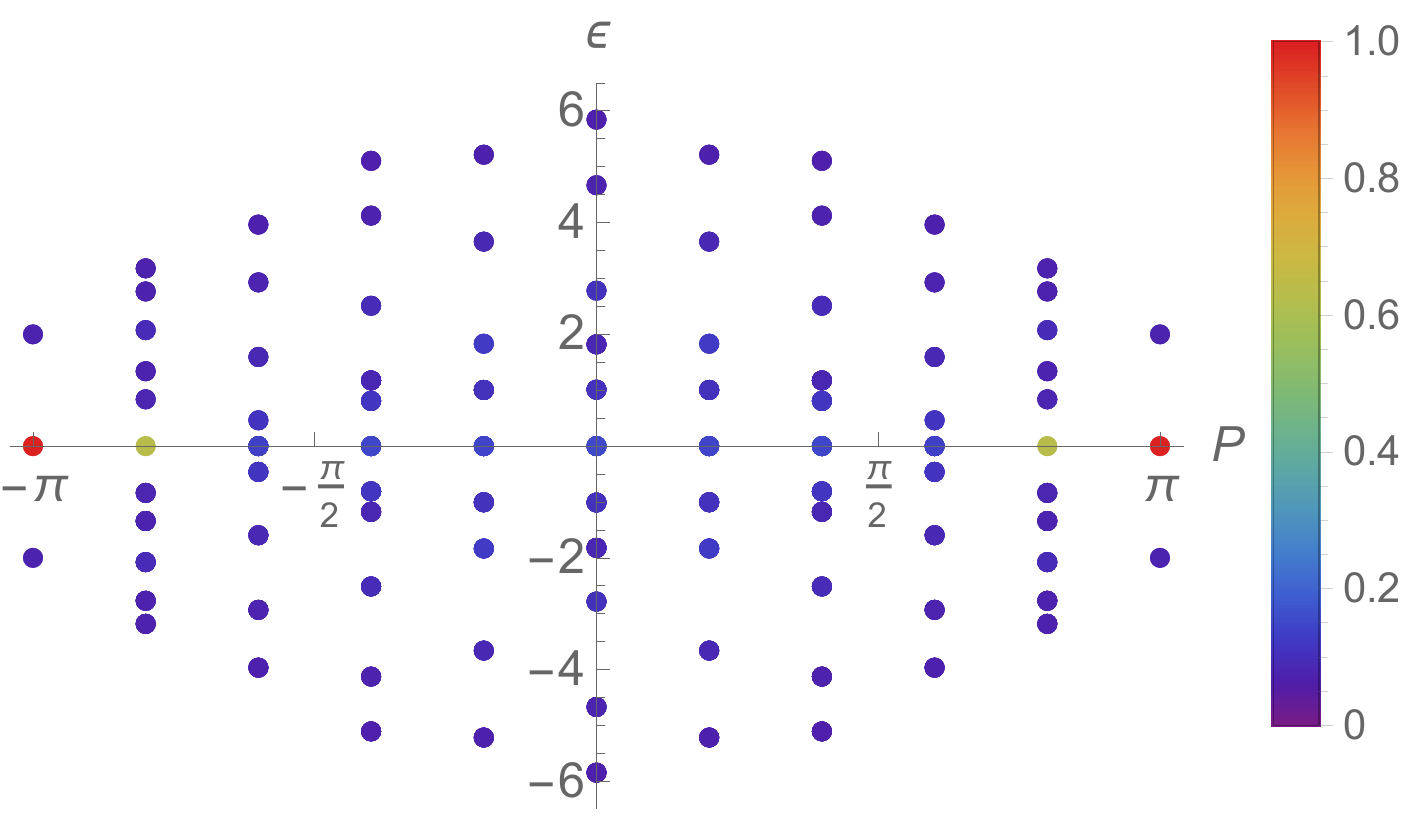}
         \caption{$\Omega/J=1$, two ``continua'' overlap, all the states in the ``continua'' including those inside of overlapped region are extended. }
     \end{subfigure}
    \hfill
     \begin{subfigure}[b]{0.8\linewidth}
         \centering
         \includegraphics[width=\linewidth]{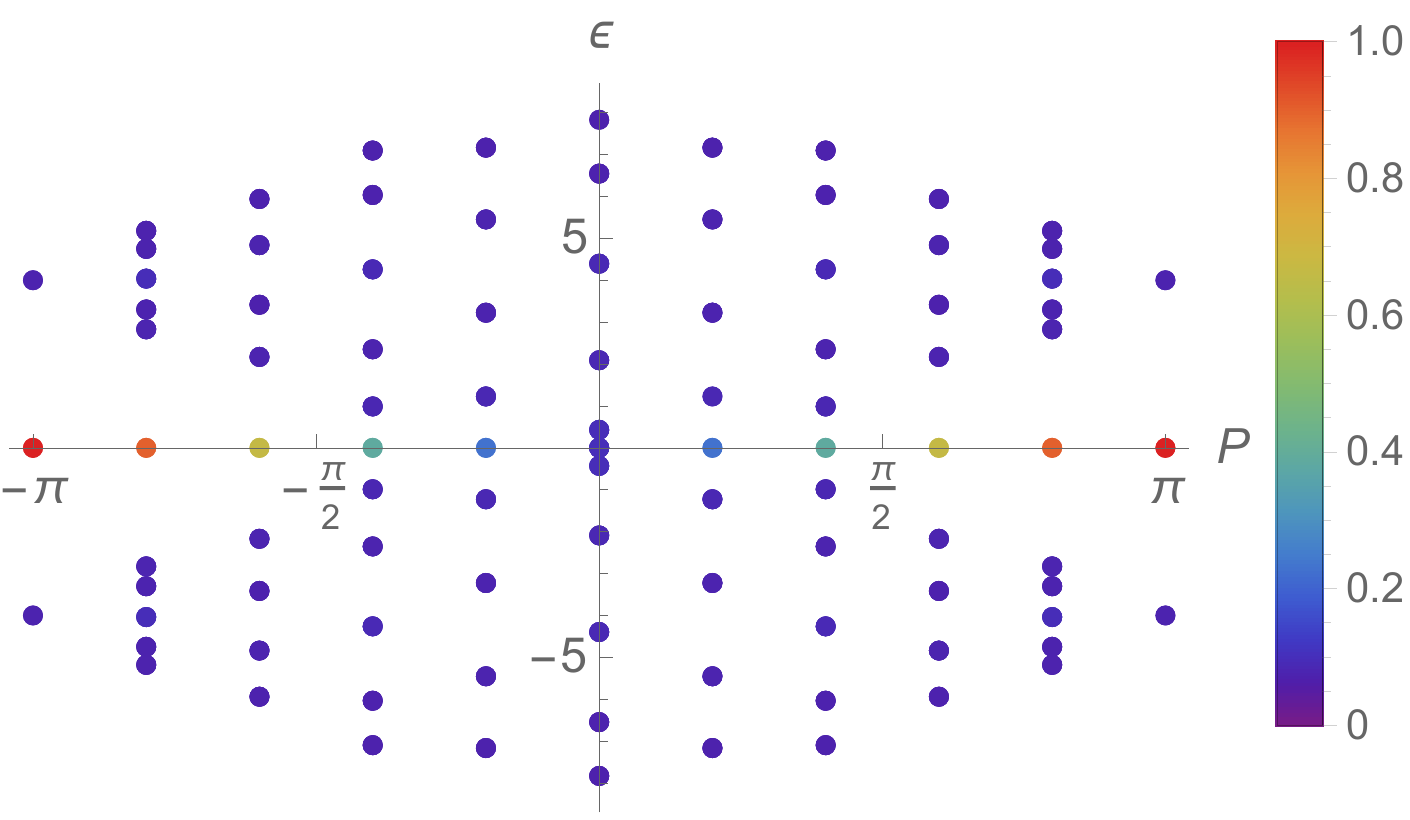}
         \caption{$\Omega/J=2$, two ``continua'' lie side by side. The doublon states become more localized when they are energetically distant from those in the ``continua''.}
     \end{subfigure}
     \hfill
      \begin{subfigure}[b]{0.8\linewidth}
        \centering
         \includegraphics[width=\linewidth]{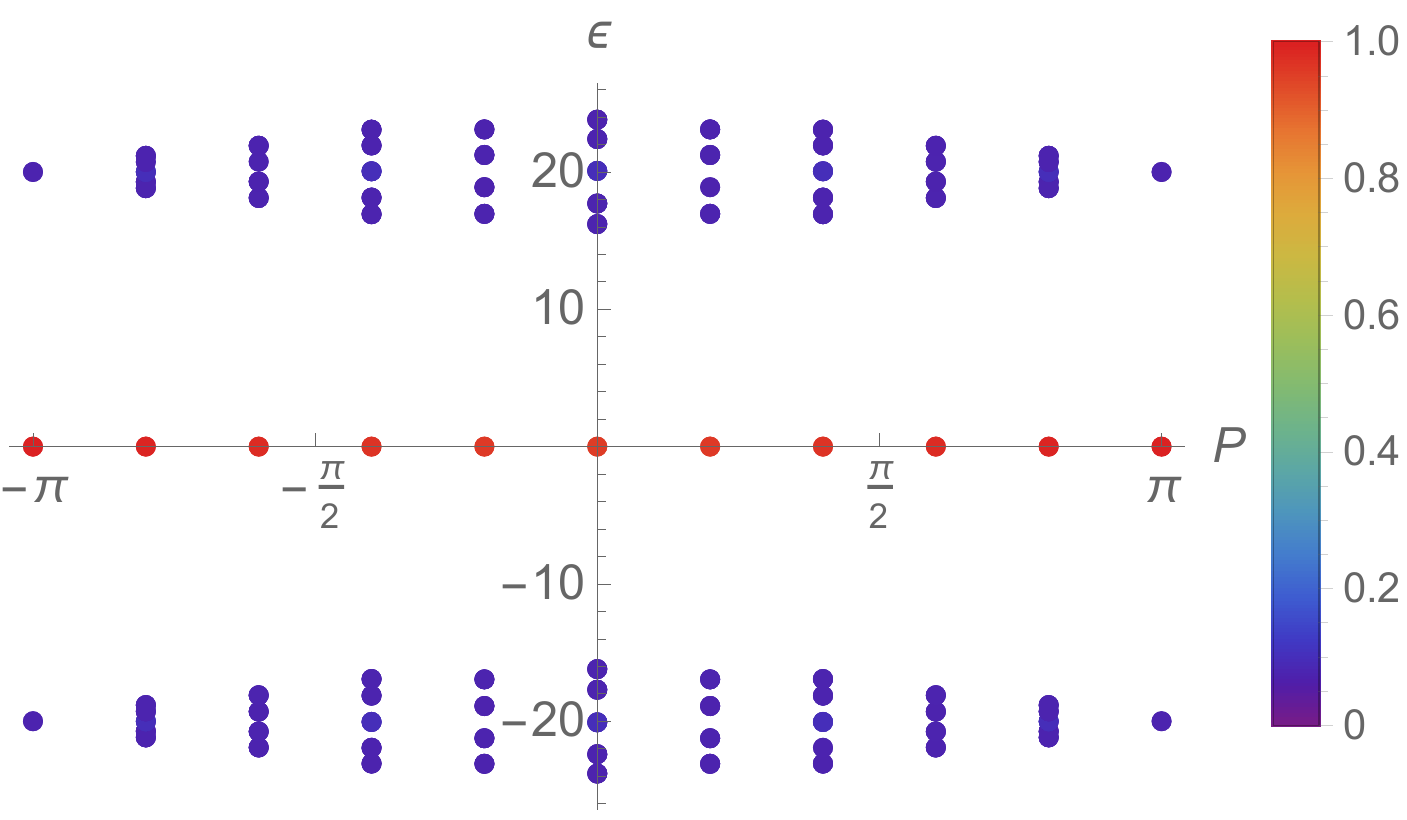}
         \caption{$\Omega/J=10$, the doublon states are almost maximally localized. Their IPR are reaching 0.1, which is the maximal value assuming translation invariance.}
     \end{subfigure}
     \caption{The spectrum ($\epsilon$) and the IPR (represented by color) of type-2 states vs. momentum $P$ when $U_1=U_2=U\rightarrow\infty$  under $\Omega/J=1,2,10$. The total site number is taken to be $N=10$.}
     \label{spectrum-2d}
\end{figure}

\medskip\noindent {\bf Inverse Participation Ratio}. We use the inverse participation ratio (IPR) to characterize and demonstrate the localization properties of all the eigenstates~\cite{Kramer_1993}. The IPR in the single particle case is defined as the integral of square of density over the space, i.e., $IPR=\sum_{i}|\psi_i|^4$. In our system we choose  two-particle spatial basis to define the IPR, which is a special case of the many-body IPR (see e.g. Ref.~\cite{2009JPhB...42l1001V}), and we obtain,
\beq
IPR=\sum_{n,m}4|A_{n,m}|^4+|B_{n,m}|^4+4|C_{n,m}|^4,
\eeq
under our previous normalization in Eq.~(\ref{eq:normalization}). In the case of $N=10$ and $U_1=U_2=\infty$, Fig.~\ref{spectrum-2d} shows the whole spectrum of type-2 states. When $\Omega/J=10$, states in the two continua are pretty much extended (IPR$<0.01$), and the doublon states are highly localized (IPR$\approx 0.1$, which is the largest IPR of states under translation symmetry).

When $\Omega/J=1$, the two continua intersect around $\epsilon=0$. The doublon states thus vanish, replaced by the more extended states in region III. These eigenstates are composed of two Choy-Haldane states with real quasi-momenta, i.e., they are essentially combinations of scattering states. Therefore their IPRs are expected to be lower than all the other states.

Therefore, we observe a hierarchy of localization in the spectrum. The most localized states are the doublons, followed  by the states composed by one scattering state and one localized state, and then the states composed by two scattering states, which are the most delocalized among the three kinds. Notice that the doublon in between two continua and the states of the third kind cannot appear simultaneously. When $\Omega$ is large, the two continua are distant, the doublon emerges in between as interaction is turned on; when $\Omega$ is small, the two continua intersect, the doublon  is replaced by the scattering states. We note that since IPR just depends on the density distribution of the system, it can in principle be measured by imaging the system~\cite{PhysRevX.9.021001}.

\medskip\noindent {\bf Entanglement}.
We also calculate the entanglement between the two species, $a$ and $b$. To do this, we define the reduced density matrix $\rho_a$ for $a$ species by tracing over the degrees of freedom in $b$-particle Hilbert space. Given that there are at most two $b$ particles, we have $\rho_a$ for a given  two-excitation state $|\psi\rangle$ defined as,
\begin{eqnarray}
\rho_a &=& {}_b\langle 0|\cdot |\psi\rangle\langle\psi|\cdot |0\rangle_{b} +\sum_i  {}_b\langle 0|b_i |\psi\rangle\langle\psi| b_i^\dagger |0\rangle_{b} \\
&& + \sum_{ij}\frac{1}{2}\,{}_b\langle 0|b_jb_i |\psi\rangle\langle\psi| b_i^\dagger b_j^\dagger |0\rangle_{b},
\end{eqnarray}
where we have used $|0\rangle_b$ to define the vacuum for $b$ particles. We expect that $\rho_a$ will be block-diagonal with three blocks contributed by the vacuum, one-particle and two-particle subspaces, respectively,
\beq
\rho_a=\rho_2\oplus\rho_1\oplus\rho_0.
\eeq

\begin{figure}[h]
     \centering
     \begin{subfigure}[b]{0.8\linewidth}
        \centering
         \includegraphics[width=\linewidth]{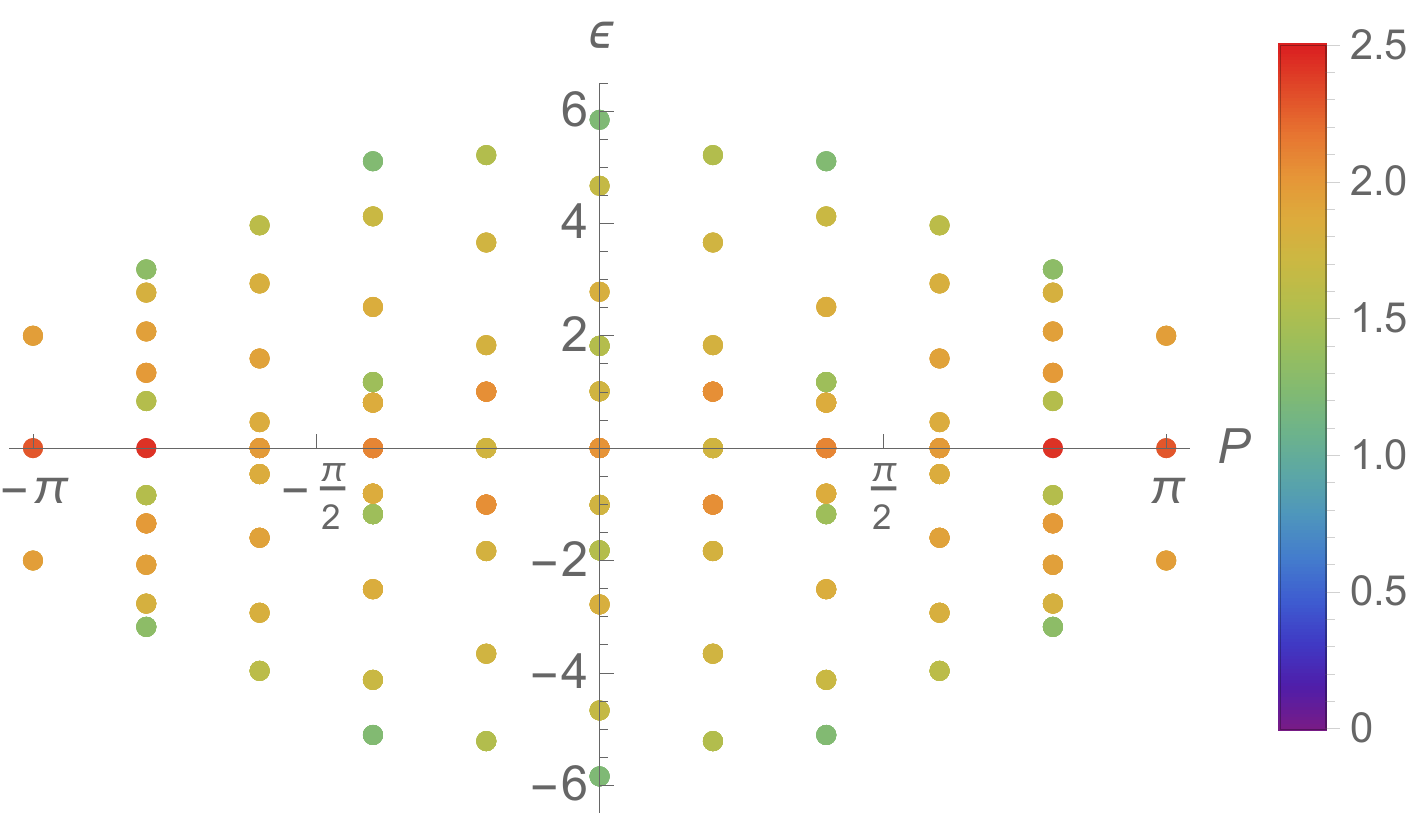}
         \caption{$\Omega/J=1$, two ``continua'' overlap. The states in the overlapped region have slightly higher entanglement between $a$ and $b$ particle. }
     \end{subfigure}
    \hfill
     \begin{subfigure}[b]{0.8\linewidth}
         \centering
         \includegraphics[width=\linewidth]{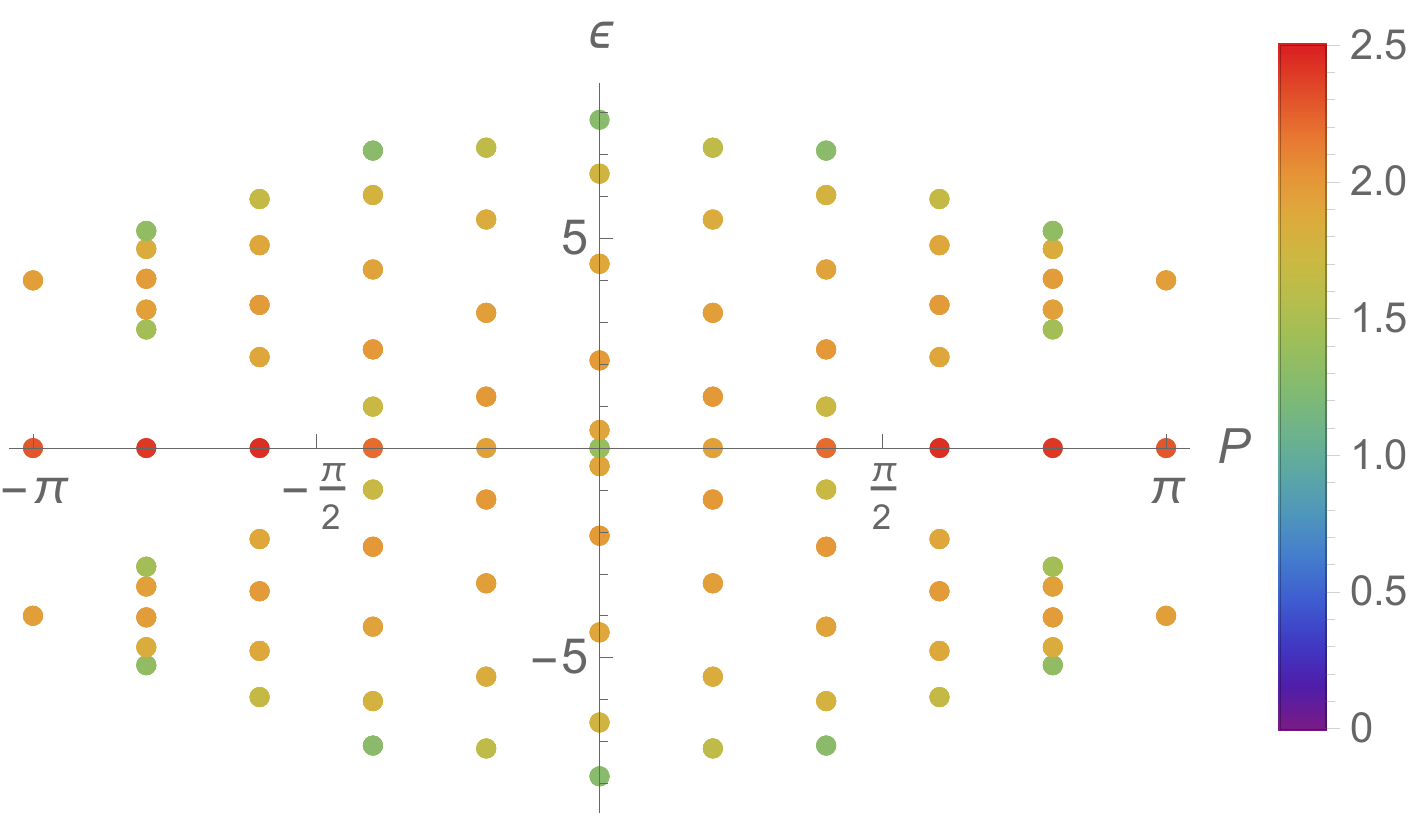}
         \caption{$\Omega/J=2$, two ``continua'' lie side by side. The doublon states have an entanglement entropy that is close to $\ln 2$.}
     \end{subfigure}
     \hfill
      \begin{subfigure}[b]{0.8\linewidth}
        \centering
         \includegraphics[width=\linewidth]{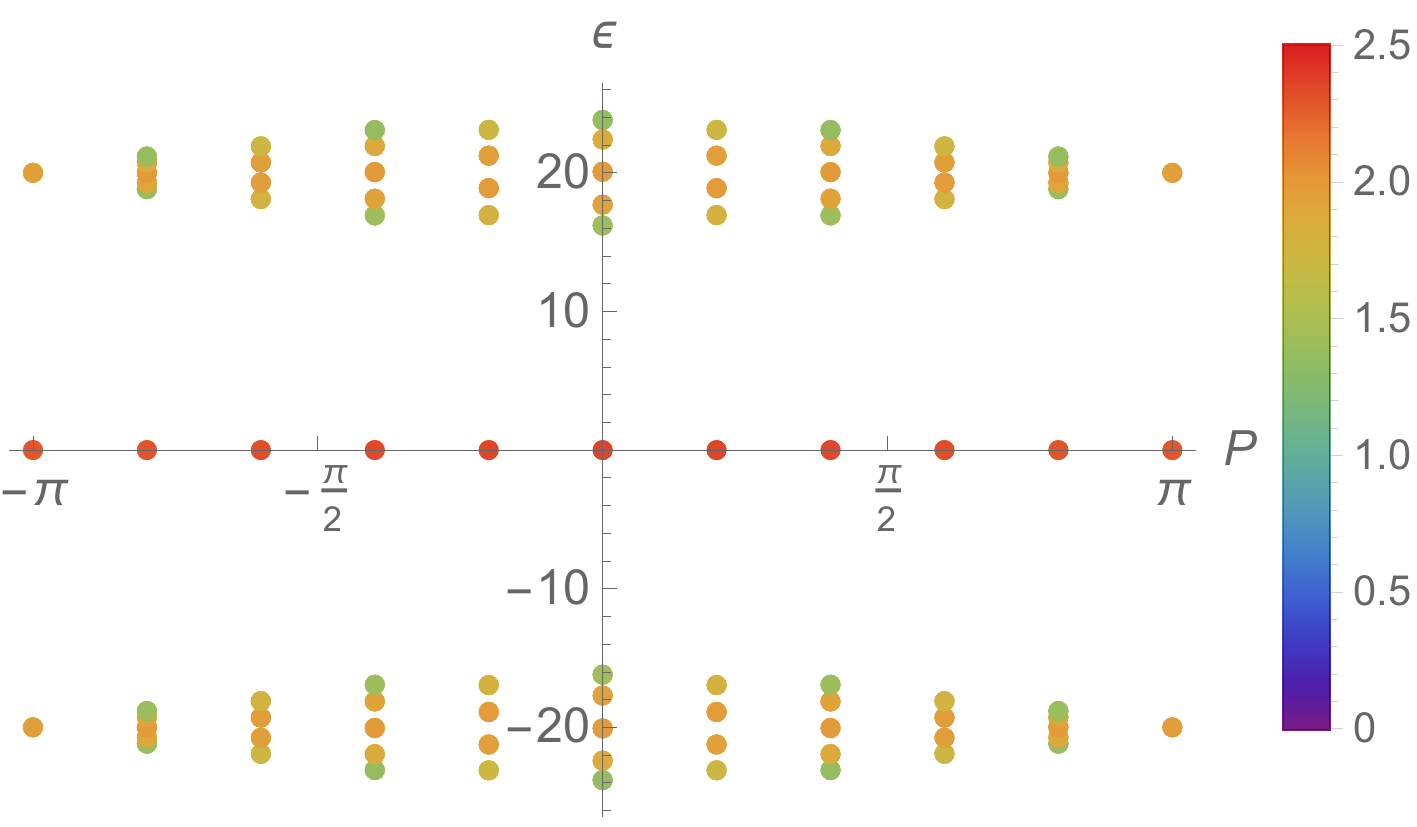}
         \caption{$\Omega/J=10$, the doublon states are almost maximal ($\approx\ln 10$) entanglement between $a$ and $b$ particle. States on the edge of each ``continuum'' have slightly lower entanglement than that of other states in the ``continuum''.}
     \end{subfigure}
     \caption{The entanglement entropy of type-2 states when $U_1=U_2=U\rightarrow\infty$ under $\Omega/J=1,2,10$. The total site number is taken to be $N=10$.}
     \label{entropy-2d}
\end{figure}

The entanglement entropy of $|\psi\rangle$ is then given by
\beq
S_{\rm vN}=-{\rm Tr}(\rho_a \ln \rho_a)=S_2+S_1+S_0
\eeq
 where $S_i\equiv-{\rm Tr}(\rho_i \ln \rho_i)$. To be more specific, we have
\begin{eqnarray}
S_2&=&-\lambda_c \ln \lambda_c
\\
S_1&=&-{\rm Tr}(BB^{\dagger}\ln BB^{\dagger})
\label{entropy}
\\
S_0&=&-\lambda_a \ln \lambda_a
\end{eqnarray}
where $\lambda_a\equiv\sum_{n,m}2|A_{n,m}|^2$, $\lambda_c\equiv\sum_{n,m}2|C_{n,m}|^2$. For type-1 eigenstates, $A=-C$, $B=0$. After imposing normalization as in Eq.~(\ref{eq:normalization}), we have $\lambda_a=\lambda_c=1/2$. Thus the entanglement entropy of type-1 eigenstates is $S=\ln 2$. We show the entanglement entropy of type-2 states in Fig.~\ref{entropy-2d}.

\medskip\noindent {\bf Site Number}.
The analysis for now in this part concentrates on the system with $N=10$ sites for simplicity and for illustration. If we increase $N$, number of states will increase for sure. Other than that, The IPR of doublon state will scale as $\sim1/N$, while IPR of states in the ``continua'' will scale as $\sim1/N^2$. The entropy of doublon states will be $\sim\ln N$, and entropy of other states will be much lower. Eventually when the thermodynamic limit is reached, states form two 4 real continua (1 in type-1, 2 in type-2, 1 in type-3).

\section{Time evolution}
\label{sec:VII}

\begin{figure}[h]
\begin{subfigure}[b]{0.49\linewidth}
     \centering
\includegraphics[width=0.99\linewidth]{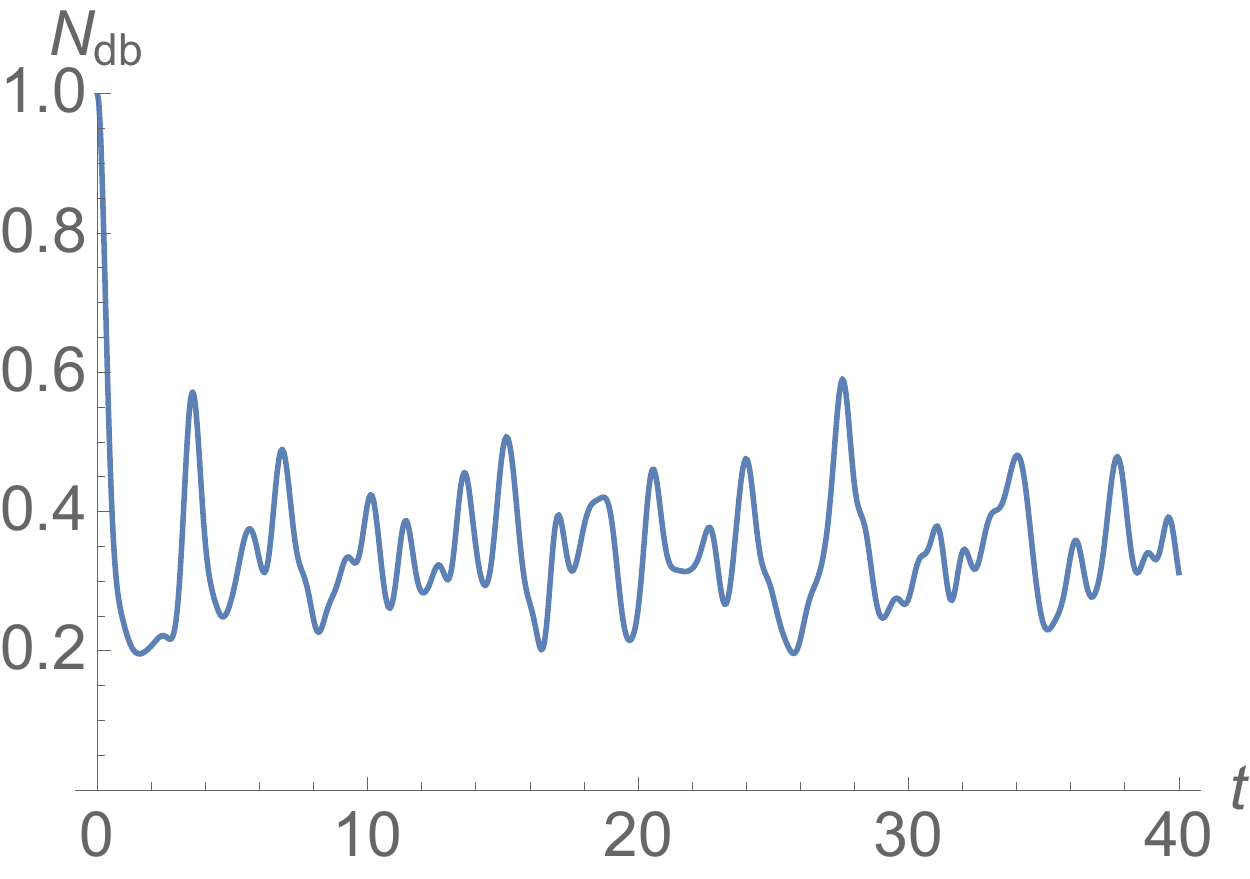}
\caption{$\Omega/J=1$, $N_{db}$ decays and oscillates drastically. }
     \end{subfigure}
    \hfill
    \begin{subfigure}[b]{0.49\linewidth}
       \centering
\includegraphics[width=0.99\linewidth]{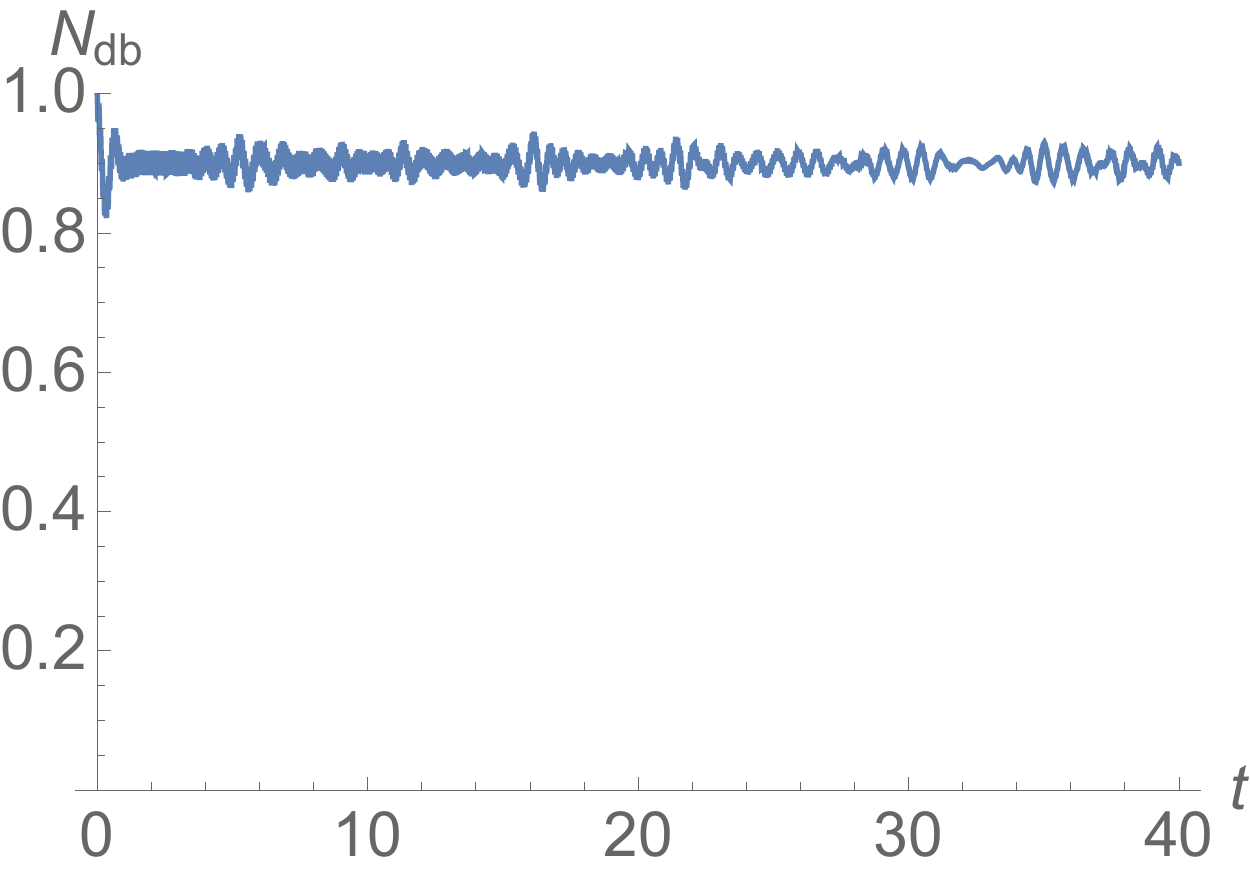}
\caption{$\Omega/J=5$, $N_{db}$ remains slightly smaller than 1.}
     \end{subfigure}
     \begin{subfigure}[b]{0.49\linewidth}
     \centering
\includegraphics[width=0.99\linewidth]{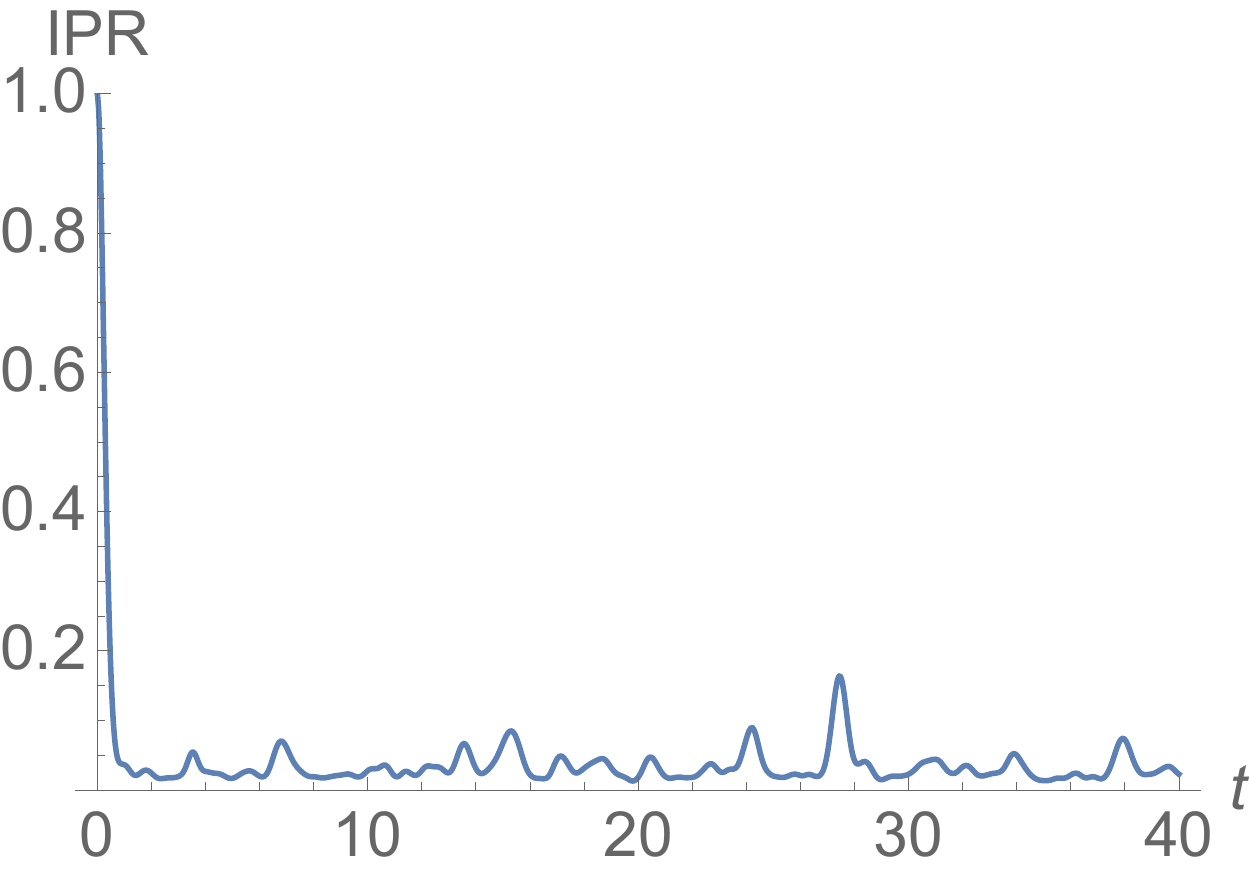}
\caption{$\Omega/J=1$, IPR decays to around 0.01 and oscillates. }
     \end{subfigure}
    \hfill
    \begin{subfigure}[b]{0.49\linewidth}
       \centering
\includegraphics[width=0.99\linewidth]{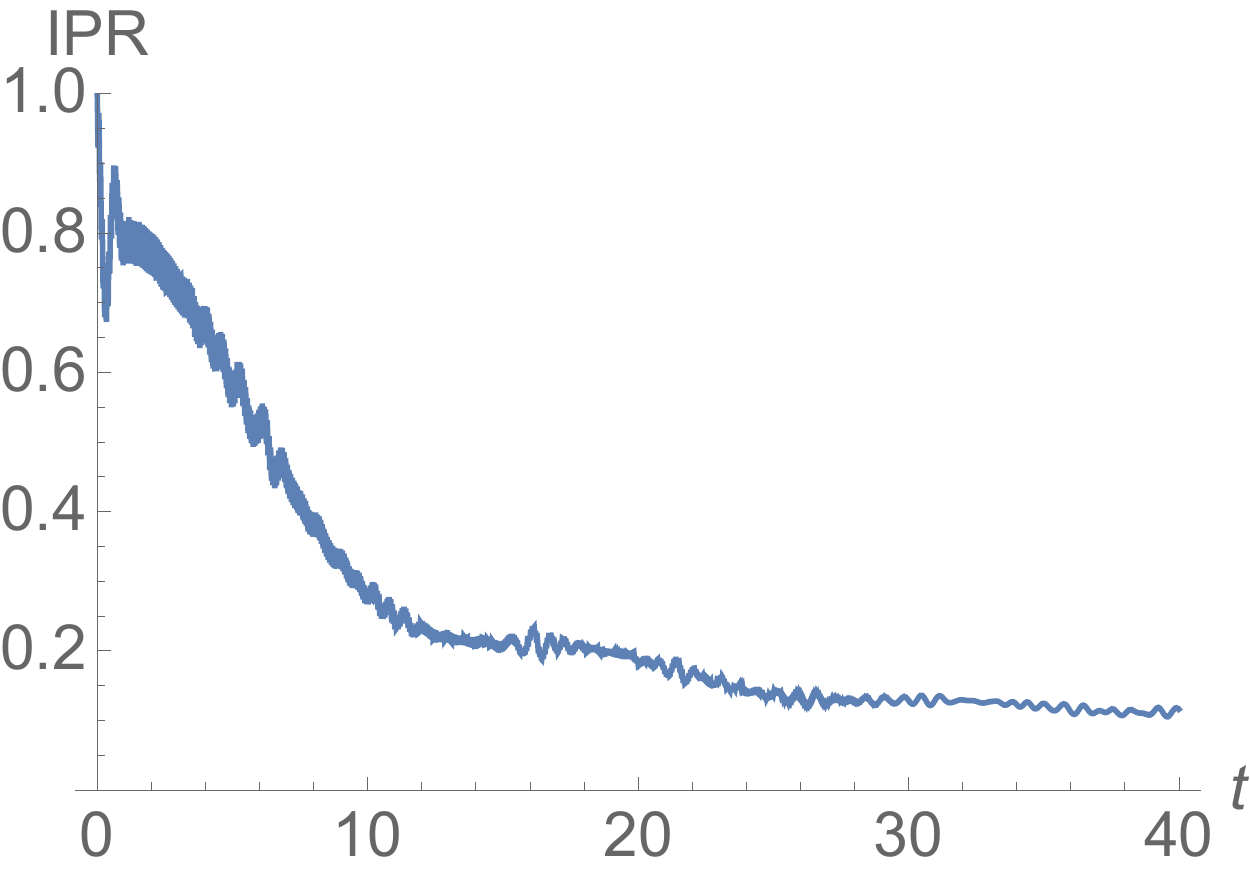}
\caption{$\Omega/J=5$, IPR decays, but has a minimum 0.1.}
     \end{subfigure}
     \caption{Time evolution of $N_{db}$ and IPR for $U/J=100$ and $N=10$ sites.}
     \label{timeevo}
\end{figure}

After analyzing detailed eigenstates in Sec.~\ref{sec:spectrum}, it is natural to ask how the system evolves from a certain initial state.  To write down the total wave function, we start with four sets of quasi-momenta which satisfy their corresponding energy equations, then using the Bethe equations and equations recombining diagonal parts, we can determine $\tilde{u}$'s, $s_{k,q}$'s and $\lambda$'s. After writing out the total complete set of basis, for every initial state (e.g. $B_{nm}=\delta_{n,0}\delta_{m,0}$), we can expand the total wave function in terms of the eigen-basis. After the expansion, we obtain the time evolution of the system easily: $\ket{\psi(t)}=\sum_i c_i e^{-iE_i t} \ket{E_i}$. This approach applies to arbitrary long time. We also remark that we also integrate the time-dependent Schr\"odinger equation numerically to obtain the dynamics, which is good for time not too long so that the numerical error does not accumulate too much.

To study the time evolution of a system that possess doublon dispersions, we set the initial state to be $\ket{\psi}=a^{\dagger}_0 b^{\dagger}_0 \ket{0}$, with two bosons of different species occupying the same position, and examine the case when $U_1=U_2=U$ is large. From the flat dispersion of the doublon state, we expect the initial state to persist in the large $\Omega$ and large $U$ limit. Therefore we examine the time evolution of $N_{db}\equiv\sum_{i}|\bra{\psi}a^{\dagger}_i b^{\dagger}_i a_i b_i\ket{\psi}|^2$, which counts only the double occupation of different species at same sites. The initial value $N_{db}=1$, as shown in Fig.~\ref{timeevo}, and it decays incompletely  and persists at a high value at late times. 

\begin{figure}[t]
\centering
\includegraphics[width=0.8\linewidth]{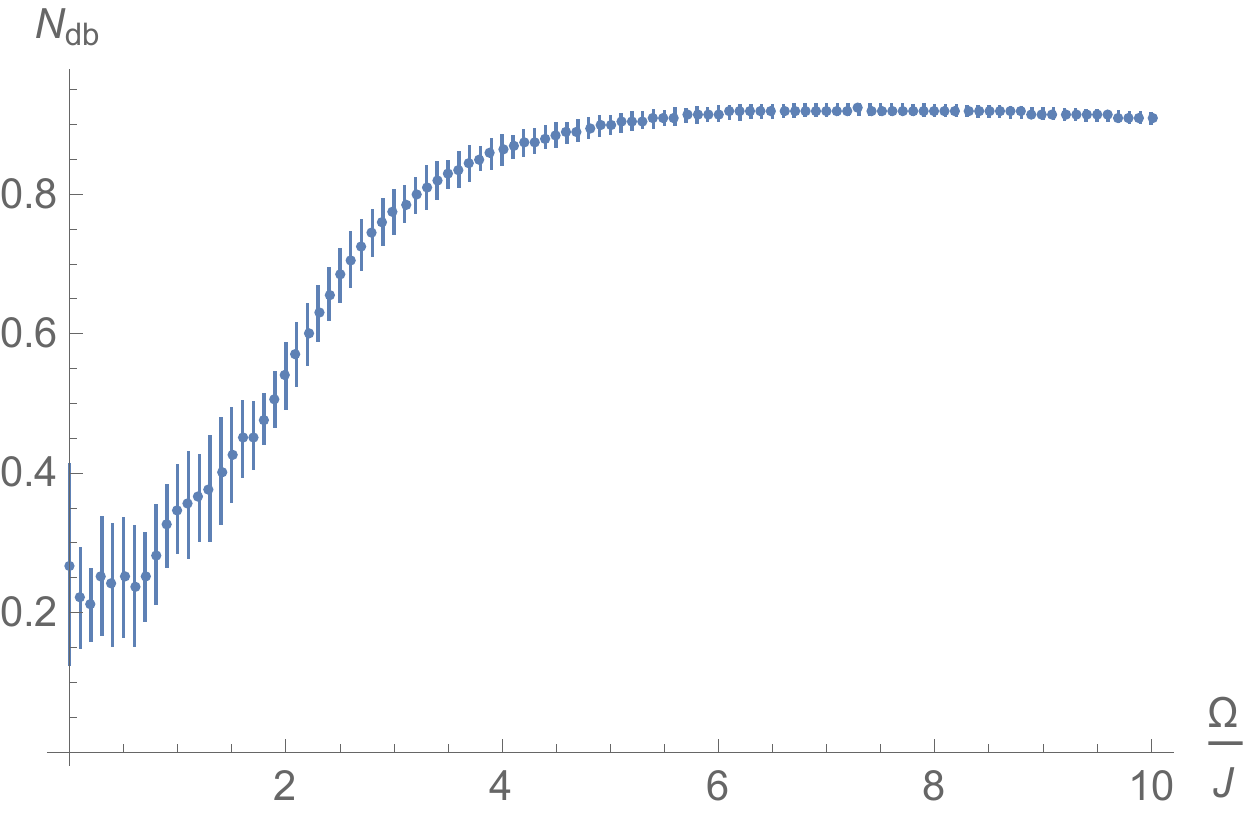}
\caption{
\label{fig:Ndb11}We show the mean values and standard deviations of $N_{db}$ between $t=30$ and $t=40$ for $U/J=100$. When $\Omega/J>2$, because of the formation of doublon, the late time mean is high, with small deviation (oscillation amplitude). When $\Omega/J$ is small, the mean is small and deviation is quite high, indicating a different late time behavior.}
\end{figure}

To study the mean and fluctuations of $N_{db}$ for a broader range of $\Omega/J$, we plot these values gathered   between time $t=30$ and $t=40$ in Fig.~\ref{fig:Ndb11}. When $\Omega/J<2$, although doublon states exist near $P=\pi$ in the momentum space, two continua overlap around $P=0$. As a result, $N_{db}$ decays drastically, giving a small mean and large deviation in the plot. When $\Omega/J>2$, once the doublon has a full band, time evolution of $N_{db}$ is dominated by the doublon.

This phenomenon is due to the localization property of doublon states, thus when $\Omega$ is small enough such that doublons do not exist, we expect a pretty different late time behavior. The difference in the late time behavior due to doublon states can be observed in the time evolution of IPR. When $\Omega/J<2$, the IPR will drastically decrease from $1$ to $0$ and oscillate slightly above it. When $\Omega/J>2$, the IPR will decrease from $1$ to somewhere above $1/N$ (which is $0.1$ in our case study) instead. 

\begin{figure}[t]
     \centering
    \begin{subfigure}[t]{0.49\linewidth}
        \centering
     \includegraphics[width=\linewidth]{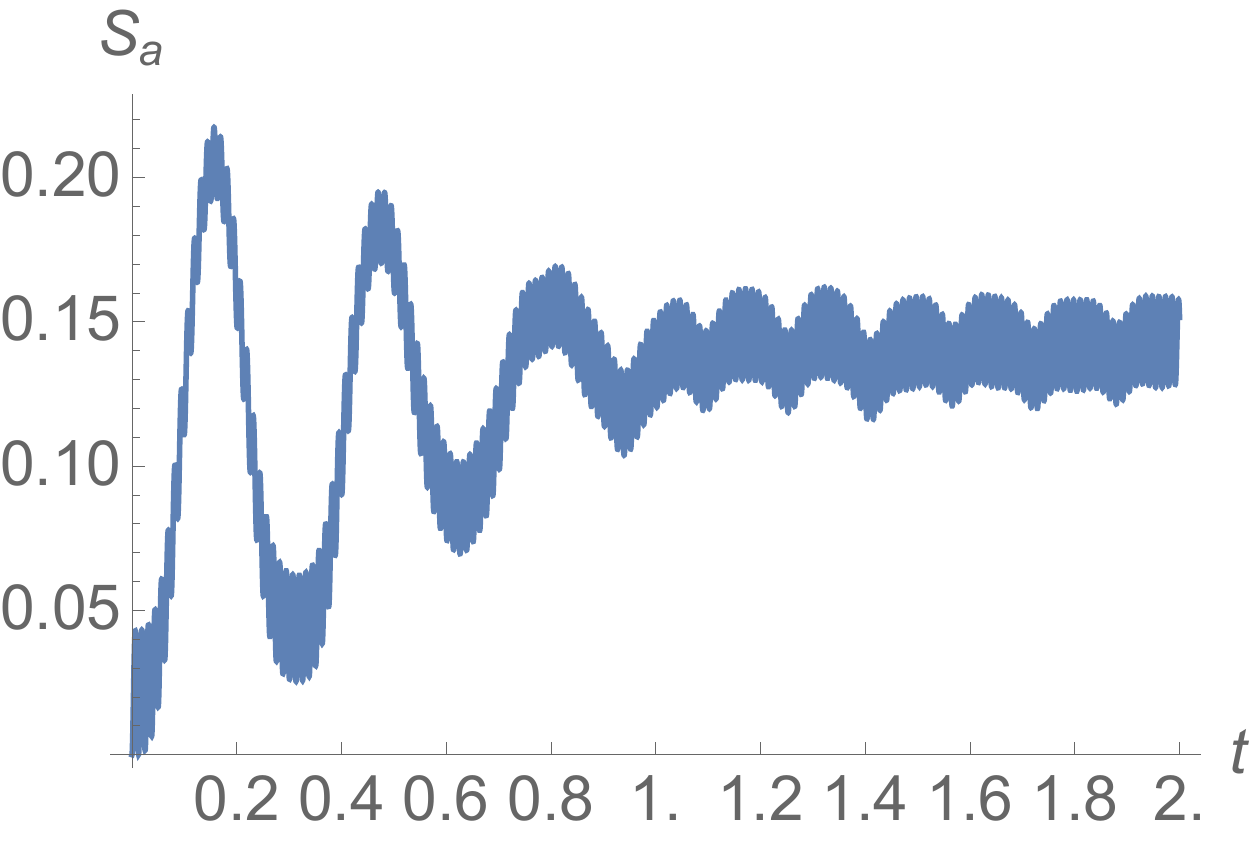}
     \caption{Entanglement from $Jt=0$ to $2$. The system build up some entropy after a short time (from $0$ to $1$).}
     \label{entbuildup}
     \end{subfigure}
     \hfill
     \begin{subfigure}[t]{0.49\linewidth}
        \centering
     \includegraphics[width=\linewidth]{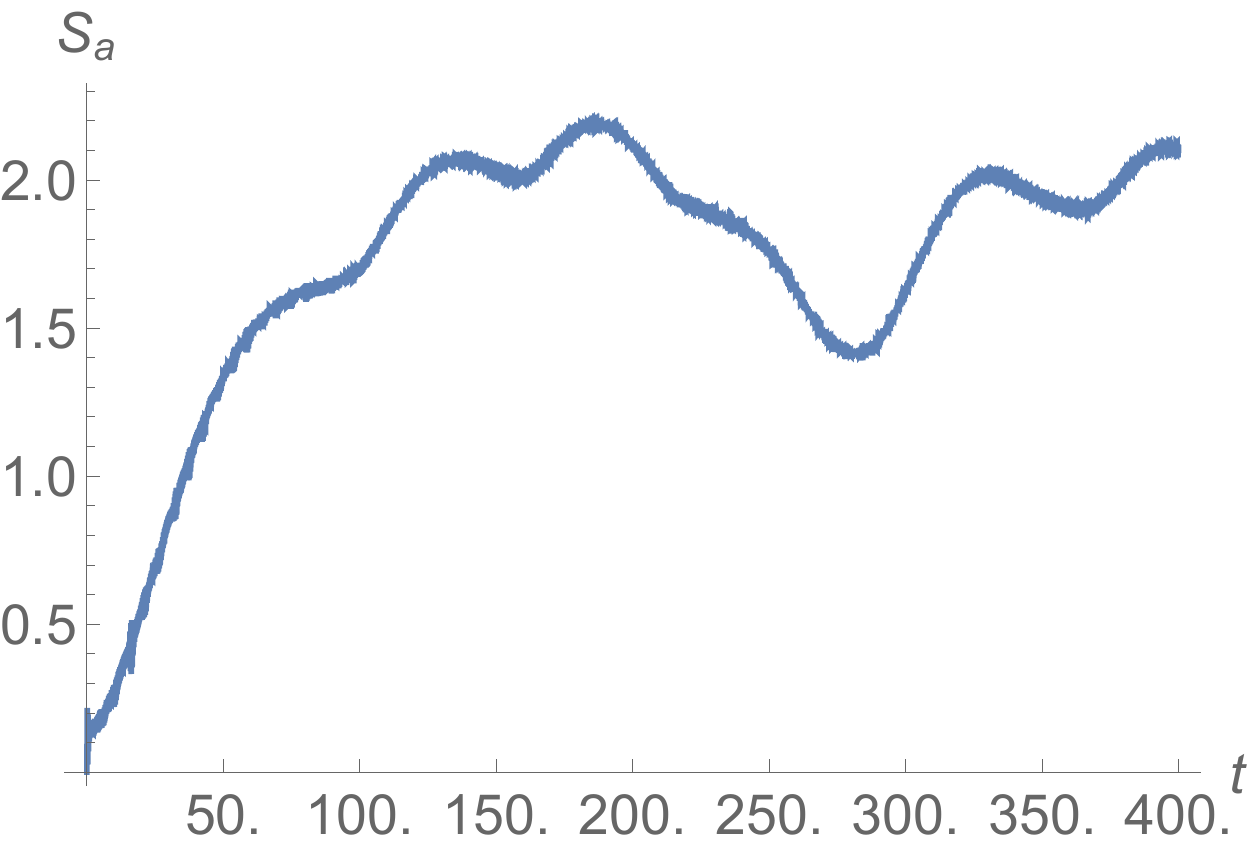}
         \caption{The time evolution of entanglement entropy from $Jt=0$ to $400$.}
         \label{entevo}
     \end{subfigure}
    \hfill
       \begin{subfigure}[t]{0.49\linewidth}
        \centering
     \includegraphics[width=\linewidth]{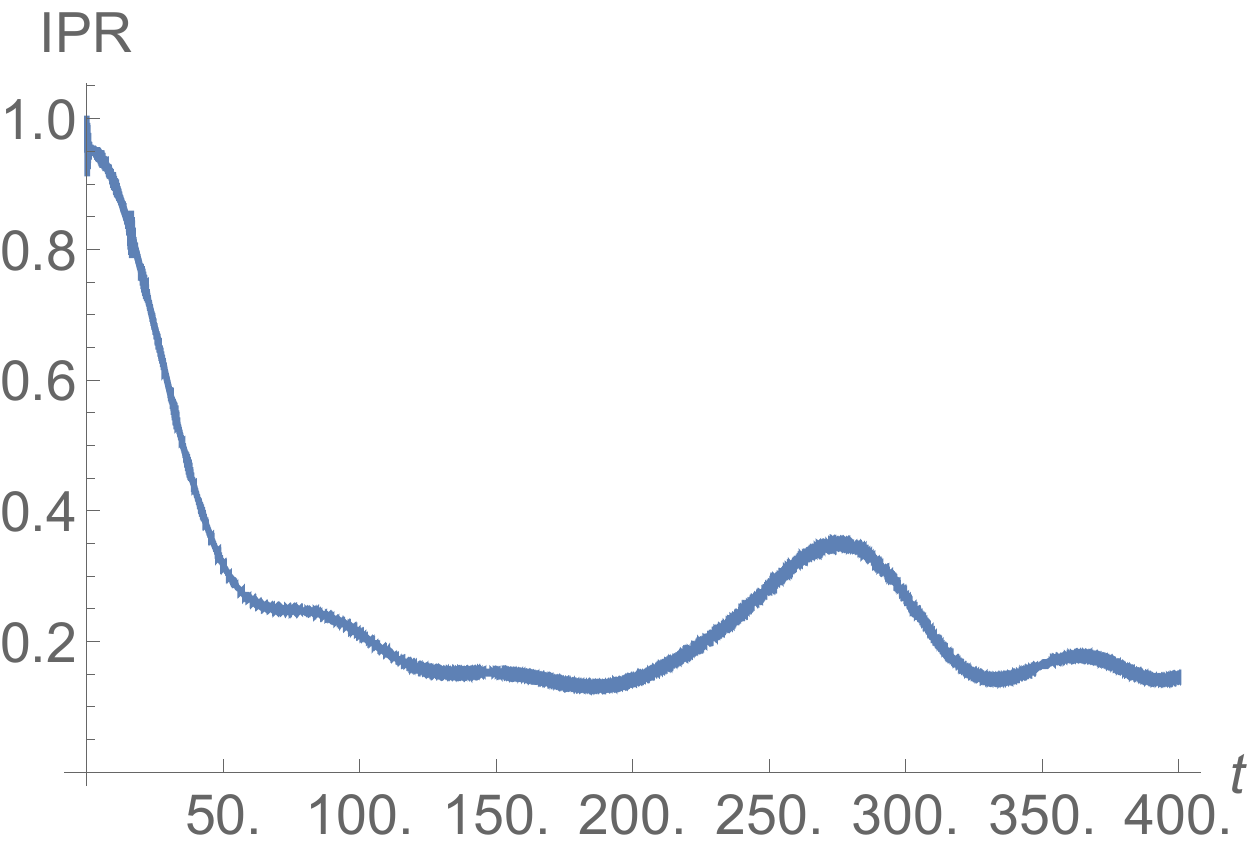}
     \caption{IPR from $Jt=0$ to $400$. It drops quickly and oscillates above $0.1$.}
     \label{iprevol}
     \end{subfigure}
      \begin{subfigure}[t]{0.49\linewidth}
         \centering
     \includegraphics[width=\linewidth]{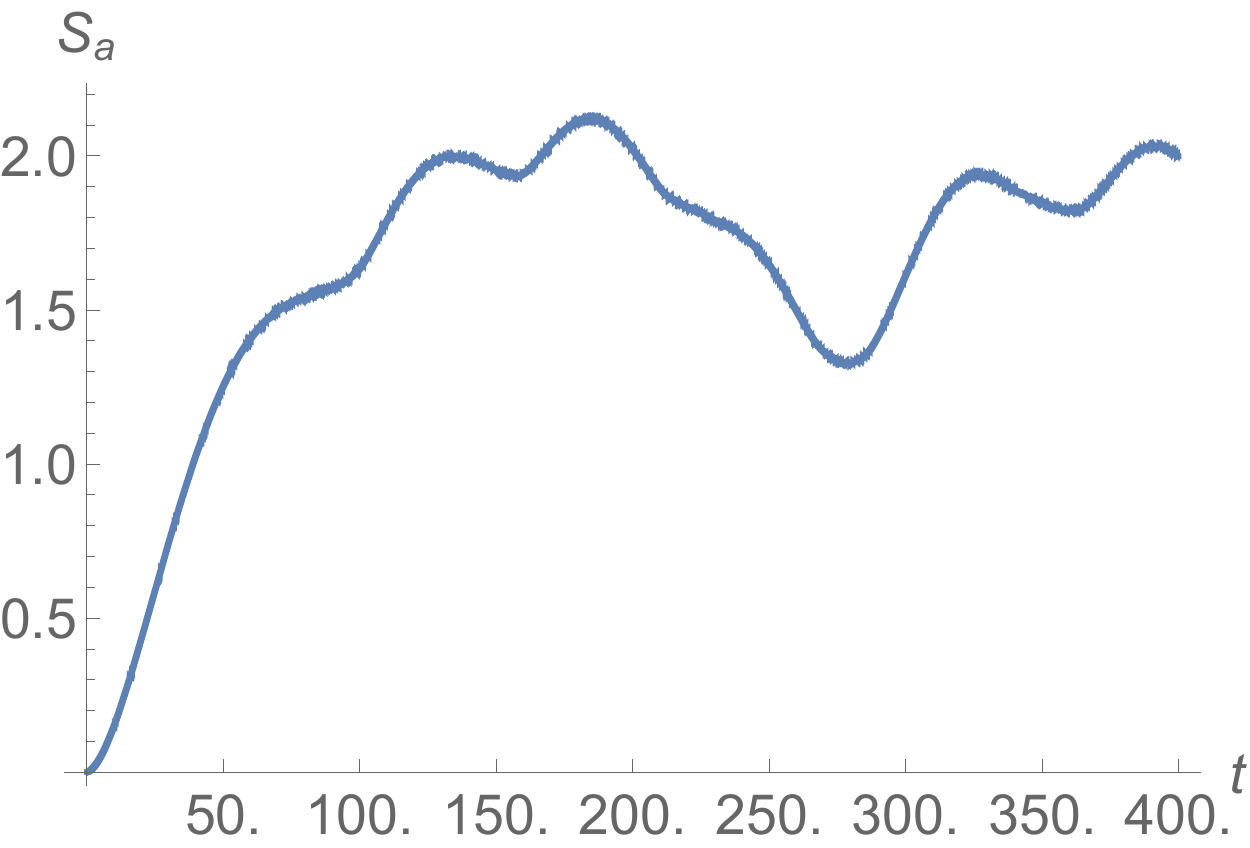}
     \caption{Entanglement between two particles in the (single-species) Bose-Hubbard model from $Jt=0$ to $400$.}
     \label{bhevo}
     \end{subfigure}
     \hfill
     \begin{subfigure}[t]{0.49\linewidth}
        \centering
     \includegraphics[width=\linewidth]{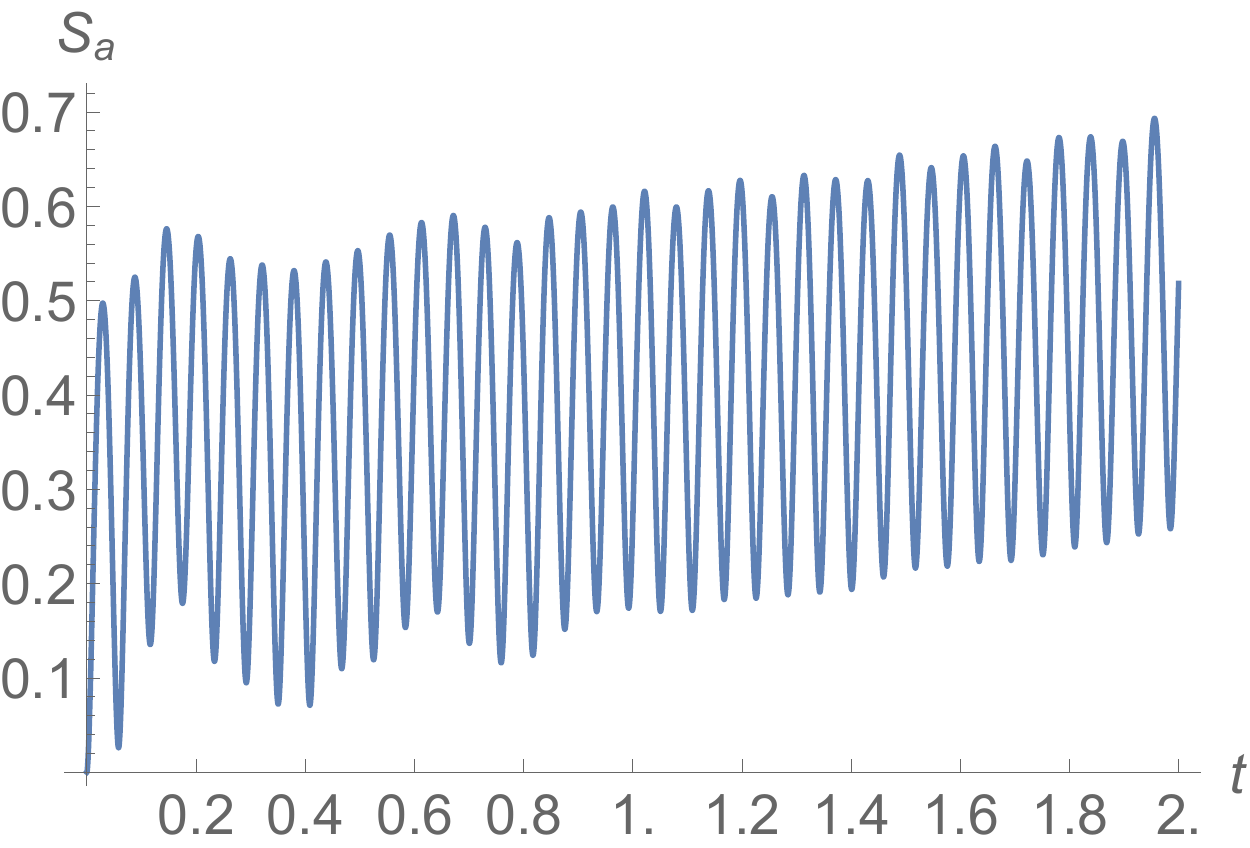}
    
     \caption{ \label{entbuildup100}Entanglement between two particles from $Jt=0$ to $2$ when $U/J=100$.}
     \end{subfigure}
     \hfill
      \begin{subfigure}[t]{0.49\linewidth}
         \centering
     \includegraphics[width=\linewidth]{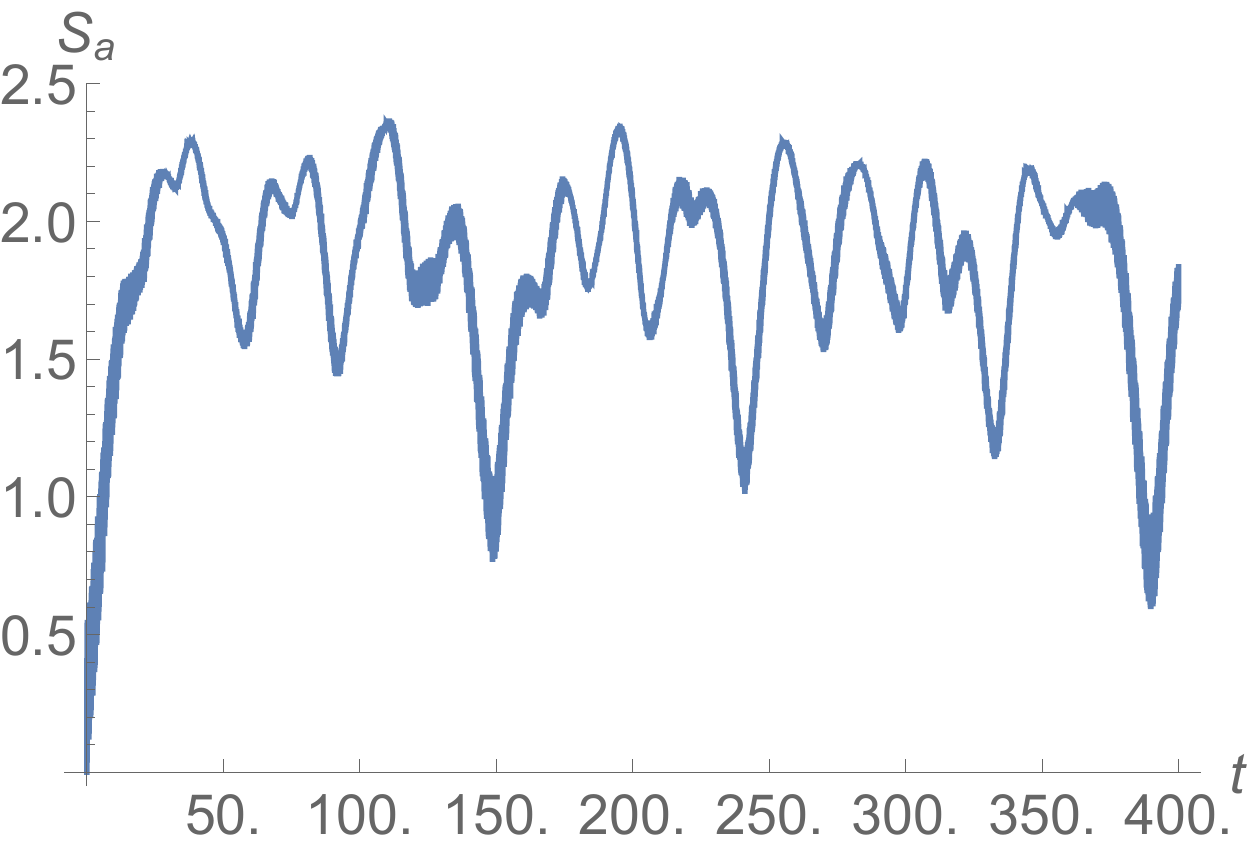}
     \caption{\label{entplot_100_400} Entanglement between two particles from $Jt=0$ to $400$ when $U/J=100$.}
     \end{subfigure}
     \hfill
     \caption{Plots a, b and c show the time evolution of coupled Bose-Hubbard model when $\Omega/J=10$, $U/J=500$. Panel (a) shows an initial buildup of entanglement in a form similar to a damped oscillator. In contrast, such an initial buildup for $U/J=100$ is weaker, as seen in (e). The panel (d) shows the time evolution of the two-particle state in the single-species Bose-Hubbard model with $U'/J=-166.7$, which is almost identical to that in (b), except the small amplitude oscillations and the initial buildup.
     Panels (e) and (f) show similar entanglement evolution for a smaller $U/J=100$, which might be more practical to realize in experiments. The oscillation pattern in (f) is 5 times faster than that in (b).}
\end{figure}

Note that when $U\rightarrow\infty$ and $\Omega\rightarrow\infty$, due to the flat doublon dispersion, the initial state is an eigenstate. Both $N_{db}$ and the IPR will remain  $1$ in the unitary evolution, and we have a standing doublon. When $U\rightarrow\infty$ and $\Omega\gg 1$, doublon dispersion is not perfectly flat, $N_{db}$ will be still close to $1$, but the IPR will drop as shown in Fig.~\ref{iprevol}.

We also study the time evolution of entanglement entropy between $a$ and $b$ particles from the same initial state of $a_0^\dagger b_0^\dagger|0\rangle$ as  above. In Fig.~\ref{entbuildup}, we show the evolution of the entanglement entropy when $U/J=500$ and $\Omega/J=10$. We  observe  very high frequency oscillation, which is easier to see in Fig.~\ref{entbuildup100} with $U/J=100$. This is due to the interference between the doublon state in region III and that in region IV. Therefore the frequency is proportional to the energy difference $\Delta\epsilon$ between the two states. When $U\gg1$ and $\Omega\gg U$, $\Delta\epsilon\propto\Omega$; when $U\gg 1$ but $U\gg\Omega$, $\Delta\epsilon\propto U$. This can be seen from comparing Fig.~\ref{entbuildup} and Fig.~\ref{entbuildup100}, where the high-frequency oscillation in the latter is slower than in the former. Moreover, the entropy varies drastically from $Jt=0$, analogous to a damped oscillation, and saturates around $Jt\approx {\rm \cal O}(1)$, with the saturation time dependent on $\Omega$. From then on after the initial buildup, the evolution is  milder.

The large time scale evolution of the entropy is shown in Fig.~\ref{entevo} with $U/J=500$. This behavior is due to the dispersion of the doublon. When $U\gg1$ and $\Omega$ is large, from Eq.~(\ref{doublon_dispersion}), the group velocity of the doublon is \beq
\frac{d\epsilon}{dP}\approx\frac{1}{U}12J^2\sin{P}.
\eeq

To demonstrate our claim regarding the behavior of entanglement, the simple way is to use the single-species Bose-Hubbard model under time evolution. In this case, when the interaction strength $U'\gg1$ (we add $'$ to $U$ to distinguish it from the interaction strengths in the coupled case), using the energy equation $\epsilon=\sqrt{U'^2+16J^2\cos^2{\frac{P}{2}}}$. The doublon group velocity is
\beq
\frac{d\epsilon}{dP}\approx-\frac{1}{U'}4J^2\sin{P}.
\eeq

After the above preparation, we can now study the entanglement entropy evolution between the two particles in the single-species Bose-Hubbard model with $U'=-\frac{U}{3}$ and the initial state being $\ket{\psi}_0=\frac{1}{\sqrt{2}}a_0^{\dagger}a_0^{\dagger}\ket{0}$. The entanglement behavior in the single-species case is expected to be similar to that in the coupled case with $U=-3U'$. 
The expression in Eq.~(\ref{entropy}) for two species motivates us to define the entanglement entropy for the single-species case as $S=-\rm Tr (AA^{\dagger}\ln AA^{\dagger})$ for the wave function of the form $\ket{\psi}=\sum_{n,m}A_{n,m}a_n^{\dagger}a_m^{\dagger}\ket{0}$. As long as $U\gg1$, we should expect a similar result in both single- and double-species cases, since the doublons in two systems have roughly the same dispersion. We demonstrate this for $U'=-500/3$ (corresponding to $U=500$) in Fig.~\ref{bhevo}, in which the evolution is essentially the same as in Fig.~\ref{entevo}, except that in the (single-species) Bose-Hubbard model there is no initial buildup process. Additionally, when $U/J$ is lowered to 100, as in Fig.~\ref{entplot_100_400} the oscillation pattern is 5 times faster than that in Fig.~\ref{entevo}.
We also see that in Fig.~\ref{entevo} and Fig.~\ref{iprevol}, when the entropy increases, IPR decreases, and vice versa. This confirms the claim that the large time scale evolution is due to the dispersion of the doublon. Notice that the IPR oscillates above $0.1$ at large times; ${\rm IPR}=1/N=0.1$ is the value of IPR when the probability density is evenly distributed in $N$ lattice sites.

\section{Conclusion}
\label{sec:conclude}
We have solved the two-particle spectrum of a generic doubly coupled Bose-Hubbard model. Our solutions include several different two-particle continua and  doublon dispersions outside of the continua. The continua are composed of states whose wave functions are superposition of Choy-Haldane states in general. In addition to these extended states, we have also obtained doublon states that are localized (in the relative coordinate). These doublons can differ from those in the single-species Bose-Hubbard model in the pattern of spatial occupation (or the adjacency feature referred earlier), such as nearest-neighbor vs. on-site. Some doublons possess energies that overlap with  a continuum, and hence they are also bound states in the continuum.  

Given that we were able to solve for all two-particle eigenstates, 
we have examined the inverse participation ratio and entanglement entropy (between the two species) for eigenstates. As doublon states are localized, they do possess a large IPR. They also have a large entanglement between the two species. We have also studied  the dynamics of simple initial two-particle states and found that the IPR and entanglement behave in opposite ways under the time evolution of the coupled Bose-Hubbard Hamiltonian. It is interesting that the behavior of  entanglement is dominated by doublons at late times, and this is confirmed by the similar  single-species BH case. The correspondence between the coupled case and the single-species case is identified by a relation in their interactions (e.g. $U=-3U'$) inferred from the doublon dispersions.
We remark that our approach can be extended to include next nearest-neighbor or longer-range hopping  and to an arbitrary number of Bose-Hubbard models that are coupled, albeit still only  one- and two-particle solutions.

Although our paper  focuses on the 2-particle subspace, the existence of localized doublons even when interaction is repulsive should change the behavior of the system drastically, including the $N_{db}$, IPR and entanglement evolution discussed in this paper. In experiments, when the doublons are localized enough (when $\Omega$ is large enough),  one could observe the doublons and effects they have in a sparse lattice. The most obvious method is to image the system using e.g. a quantum gas microscope~\cite{bakr2009quantum}. When $U_1=U_2\gg1$, the group velocity of doublons approaches zero, which means that if one carefully specifies the initial state as several doublons lying in a sparse lattice, all the particles will essentially freeze at their initial positions.  One could vary the interactions and observe different dynamics. In a different approach, one could also use modulation spectroscopy and time of flight techniques to detect the presence of  doublons, similar to the single-species case in ~\cite{winkler2006repulsively}.

Finally, a digital quantum computer may also be used to simulate the system. When $U_1=U_2=\infty$, the $a$ and $b$ particles are hard core bosons. So the two-particle wave function can be simulated by $a$ and $b$ types of qubits, with coupling on site. However, when the intra-species interaction is finite, the two particle wave function should be simulated by 2 types of qutrits instead of qubits, as there are three possible occupations $|0\rangle$, $|1\rangle$, and $|2\rangle$ of the same species. Their dynamics can be studied in principle via a Trotter-Suzuki decomposition of the time evolution operator into a sequence of quantum gates. The IPR, entanglement (perhaps using the Renyi-2  instead of the von Neumann entropy)  and diagonal occupation can all be studied by measurements. At the moment, qubit digital quantum computers look more promising for hard-core bosons, but some recent experiments have begun to explore  qutrits~\cite{PhysRevX.11.021010}, which might be used to simulate general two-excitation states in the coupled Bose-Hubbard model.

\begin{acknowledgments}
This work was  supported by National Science Foundation Grant Nos. PHY-1915165 (Y.L. and T.-C. W.) and PHY-1912546 (D.S.) as well as an SBU seed grant (Y.L, D.S. and T.-C. W.) on the initial design of the project.
\end{acknowledgments}
\begin{widetext}
\appendix
\section{Momentum space Schrodinger equation}
\label{app:momentum}
\beq
(\epsilon-\omega_p-\omega_q) A_{p,q}&= \frac{U_1}{N}\sum_{p+q=P}A_{p,q}+\frac{\Omega}{2}(B_{p,q}+B_{q,p}),
\\
(\epsilon-\omega_p-\omega'_{q}) B_{p,q}&=2\Omega A_{p,q} + 2\Omega C_{p,q},
\\
(\epsilon-\omega'_p-\omega'_q)C_{p,q}&=\frac{U_2}{N}\sum_{p+q=P}C_{p,q}+\frac{\Omega}{2}(B_{p,q}+B_{q,p}).
\eeq
Substituting the third one into the second, we have
\beq
(\epsilon-\omega_p-\omega_q) A_{p,q}&= \frac{U_1 A}{N}+\frac{\Omega}{2}(B_{p,q}+B_{q,p}),
\\
2\Omega A_{p,q}&=(\epsilon-\omega_p-\omega'_{q}) B_{p,q}-\Omega^2\frac{ B_{p,q}+B_{q,p}}{\epsilon-\omega'_p-\omega'_{q}}+\frac{2\Omega U_2 C}{N(\epsilon-\omega'_p-\omega'_{q})}.
\label{a-b-new}
\eeq
Substituting the second one into the first one, we have
\beq
(\epsilon-\omega_p-\omega'_{q}) B_{p,q}-\Omega^2(\frac{1}{\epsilon-\omega_p-\omega_{q}}+\frac{1}{\epsilon-\omega'_p-\omega'_q})(B_{p,q}+B_{q,p})
\\
=\frac{2\Omega U_1 A}{N}\frac{1}{\epsilon-\omega_p-\omega_q}-\frac{2\Omega U_2 C}{N}\frac{1}{\epsilon-\omega'_p-\omega'_q}.
\label{Bpq}
\eeq
From equation~(\ref{a-b-new}) we could further have
\beq
 \frac{\epsilon-\omega_p-\omega'_{q}}{2\Omega}B_{p,q}=\frac{2\epsilon-\omega_p-\omega_{q}-\omega'_p-\omega'_q}{\epsilon-\omega'_p-\omega'_q}A_{p,q}-\frac{ U_1 A}{N}\frac{1}{\epsilon-\omega'_p-\omega'_q}-\frac{ U_2 C}{N}\frac{1}{\epsilon-\omega'_p-\omega'_q}.
\eeq
Substituting this into equation(\ref{Bpq}) and taking advantage of the fact that $A_{p,q}=A_{q,p}$, we arrive at
\beq
A_{p,q}&=\frac{1}{(\epsilon-\omega_p-\omega_q-\eta_{pq})(\epsilon-\omega'_p-\omega'_q-\eta_{pq})-\eta_{pq}^2} \Big( (\epsilon-\omega'_p-\omega'_q-\eta_{pq})\frac{U_1 A}{N}+\eta_{pq}\frac{U_2 C}{N}\Big),
\eeq
where $\eta_{pq}\equiv\Omega^2(\frac{1}{\epsilon-\omega_p -\omega'_q}+\frac{1}{\epsilon-\omega_q -\omega'_p})$. Similarly for $C_{p,q}$,
\beq
C_{p,q}&=\frac{1}{(\epsilon-\omega_p-\omega_q-\eta_{pq})(\epsilon-\omega'_p-\omega'_q-\eta_{pq})-\eta_{pq}^2} \Big( \eta_{pq}\frac{U_1 A}{N}+(\epsilon-\omega_p-\omega_q-\eta_{pq})\frac{U_2 C}{N}\Big).
\label{ac}
\eeq
This essentially forms a matrix equation
\beq
\begin{pmatrix}
A_p \\
C_p
\end{pmatrix}
=
M(p)
\begin{pmatrix}
\frac{A}{N}
\\
\frac{C}{N}
\end{pmatrix}.
\eeq
Since $\sum_p A_p=A$ and $\sum_p C_p=C$, we have an equation for the matrix $M$ which essentially becomes the energy equation,
\beq
\det ( \sum_p M(p) -\mathbf{1})=0,
\eeq
or in the thermodynamic limit, the sum becomes an integral
\beq
\det (\int_0^{2\pi}\frac{dp}{2\pi}M(p)-\mathbf{1})=0.
\eeq

As an example, when $U_2=0$ and $J_2=0$, the matrix equation reduces to ordinary equation, then the energy equation is 
\beq
\label{appeq:epsilonU}
\sum_{p+q=P}\frac{ U}{N}\frac{(\epsilon-\omega_p-\omega_{q})-(\frac{\Omega^2}{\epsilon-\Delta-\omega_{q}}+\frac{\Omega^2}{\epsilon-\Delta-\omega_{p}})}{(\epsilon-2\Delta)(\epsilon-\omega_p-\omega_{q})-(2\epsilon-2\Delta-\omega_p-\omega_q)(\frac{\Omega^2}{\epsilon-\Delta-\omega_{q}}+\frac{\Omega^2}{\epsilon-\Delta-\omega_{p}})}=1.
\eeq
In the limit of $N\rightarrow\infty$, this sum turns into an integral,
\beq
\int_0^{2\pi}dp\frac{(\epsilon-\omega_p-\omega_{q})-(\frac{\Omega^2}{\epsilon-\Delta-\omega_{q}}+\frac{\Omega^2}{\epsilon-\Delta-\omega_{p}})}{(\epsilon-2\Delta)(\epsilon-\omega_p-\omega_{q})-(2\epsilon-2\Delta-\omega_p-\omega_q)(\frac{\Omega^2}{\epsilon-\Delta-\omega_{q}}+\frac{\Omega^2}{\epsilon-\Delta-\omega_{p}})}\Big|_{q=P-p}=\frac{2\pi}{U}
\eeq
From this equation, one can in principle solve for the two-excitation energy $\epsilon$. In fact, we can use this in the opposite direction, i.e., by fixing an $\epsilon$ and performing the integration (e.g. numerically) to obtain the corresponding interaction $U$. This allows us to obtain the relation between $\epsilon$ and $U$, in particular, for the doublons.

\section{Inter-species interaction}
\label{sec:VI}

So far we have not included interaction between the two species of atoms. If we have inter-species interaction, the Schr\"odinger equations  become
\begin{eqnarray}
& \epsilon A + T_1 A + A T_1 -U_1 D_A = \frac{\Omega}{2} (B+B^T),
\\
& \epsilon B+ T_1 B +B T_2 - U_3 D_B=2\Omega (A+C),
\\
& \epsilon C + T_2 C + C T_2 -U_2 D_C= \frac{\Omega}{2} (B+B^T).
\end{eqnarray}

The two-excitation states can still be solved, because the energy equations from Eq.~(\ref{energyeq}) still hold. Therefore, the wave functions are still combinations of four different Choy-Haldane states, just with different weights $\lambda$'s. Since the second equation has one extra term when inter-species interaction is non-vanishing, the equations determining $\lambda$'s become slightly modified,
\begin{eqnarray}
&&\sum_{i=1}^4 J_1 \Tilde{u}_i \lambda_i (1+s_i)=U_1 \sum_{i=1}^4 \lambda_i (1+s_i),
\\
&&\sum_{i=1}^4 J_2 \Tilde{u}_i \lambda'_i (1+s_i)=U_2 \sum_{i=1}^4 \lambda'_i (1+s_i),
\\
%&\begin{aligned}
&&\sum_{i=1}^4 \lambda_i (\epsilon-\omega_{k_i}-\omega_{q_i})\big((J_1+J_2) \Tilde{u}_i  (1+s_i)  \nonumber
\\
&& \quad - (J_1-J_2)2\mathrm{i}(\sin{k_i}+\sin{q_i})  \kappa_i (1-s_i)  \big) \nonumber
\\
&&\quad=2\Omega U_3\sum_{i=0}^4 \lambda_i (\epsilon-\omega_{k_i}-\omega_{q_i}) (1+s_i),
%\end{aligned}
\\
&&\sum_{i=1}^4 \kappa_i \lambda_i (\epsilon-\omega_{k_i}-\omega_{q_i}) (1-s_i)=0.
\end{eqnarray}

\smallskip With these equations, we now discuss a few limits.

\noindent (i) When $U_1=U$, $U_2=0$, $J_1=J_2=J$, these equations are simplified, $\lambda'_{1,2}=\lambda_{1,2}$, $\lambda'_3=-\lambda_3$. We let $u_i\equiv U_i/J$,
\begin{eqnarray}
&&\sum_{i=1}^3\lambda_i (1+s_i) \tilde{u}_i=u \sum_{i=1}^3 \lambda_i (1+s_i),
\\
&&\sum_{i=1}^3\lambda'_i (1+s_i) \tilde{u}_i=u_2\sum_{i=1}^3 \lambda'_i (1+s_i)=0,
\\
&&\sum_{i=1}^2(-1)^{i}\lambda_i (1+s_i) \tilde{u}_i
%-\lambda_2 (1+s_2) \tilde{u}_2%
=u_3(\sum_{i=1}^2(-1)^{i}\lambda_i (1+s_i)).
\end{eqnarray}

\smallskip
\noindent (ii) When $U_1=U_2=U$, $J_1=J_2=J$, these equations become,
\begin{eqnarray}
&&\lambda_1 (1+s_1) \tilde{u}_1+\lambda_2 (1+s_2) \tilde{u}_2\nonumber\\&&\quad=u ( \lambda_1 (1+s_1)+ \lambda_2 (1+s_2)),
\\
&&\lambda_1 (1+s_1) \tilde{u}_1-\lambda_2 (1+s_2) \tilde{u}_2\nonumber\\&&\quad=u_3(\lambda_1 (1+s_1)-\lambda_2(1+s_2)),
\\
&&\tilde{u}_3=u_3.
\end{eqnarray}

\smallskip
 It turns out that when $U_3\ne0$, there are the same three types of solutions as we have seen in Sec.~\ref{sec:spectrum}: 1) $A=-C=HC_0$, $B=0$. 2) $A=C=\lambda_1 HC_1+\lambda_2 HC_2$. 3) $B=-B^T$, $A=C=0$. Turning on inter-species interaction $U_3$, the first and third types of states stay the same. For the second types of states, they are still combinations of the same two Choy-Haldane states, except that $\lambda_{1,2}$ are different now. We can further take the case when $U_1=U_2=U_3=U$, then the resulting solutions of the second type are just $\tilde{u}_1=u,~\lambda_2=0$ and $\tilde{u}_2=u,~\lambda_1=0$. This means that when all the inter- and intra-species interactions are the same, the two-particle states are simply Choy-Haldane states (and anti-symmetric states): $\ket{\psi}=\ket{HC_i}$, momenta of which satisfy energy equations: $\epsilon=\omega_k+\omega_q+c$, $c=0,\pm2\Omega$.

\end{widetext}

\bibliography{Refs}% Produces the bibliography via BibTeX.

\end{document}